\documentclass[showpacs,nofootinbib]{revtex4}

\usepackage{graphicx}
\usepackage{dcolumn}
\usepackage{bm}
\usepackage{epsfig}
\usepackage{amssymb}
\usepackage{amsmath}

\usepackage{float}

\begin{document}

\preprint{APS/123-QED}

\title{ Semileptonic meson decays in point-form relativistic quantum mechanics: 
unambiguous extraction of weak form factors  }

\author{  Mar\'ia G\'omez-Rocha  }
\email{  maria.gomez-rocha@uni-graz.at  }
\affiliation{  University of Graz, 
               Institute of Physics, 
               NAWI Graz, A-8010 Graz, 
               Austria                   }

\date{\today}

\begin{abstract}

The point-form version of the Bakamjian-Thomas 
construction is applied to the description of 
several semileptonic decays of mesons. 
Weak form factors are extracted without 
ambiguity for pseudoscalar-to-pseudoscalar 
as well as for pseudoscalar-to-vector 
transitions of mesons from the most general
covariant decomposition of the weak current. 
No manifestation of cluster-separability 
violation appears in the form of non-physical 
contributions to the structure of such a current,  
in contrast to what happens in the electromagnetic case. 
Moreover, no frame dependence is observed 
when we extract the form factors from the 
most general covariant decomposition of 
the current, which contrasts with analogous 
front-form calculations that involve vector 
mesons in the transition. We present our 
results for heavy-light meson decays, 
i.e. $B\to D$, as well as for $B$ and $D$ 
mesons decaying into $\pi$, $\rho$ and 
$K^{(*)}$ and perform a numerical comparison 
with the analogous front-form approach. 
Differences between point and front forms 
that are not seen in the heavy-quark limit 
of $q\bar Q$-systems appear. These differences 
are attributed to the different role that 
the non-valence contributions play in the 
description of hadronic reactions in each form. 
It is argued how contributions from missing 
$Z$-graphs can be estimated. 

\end{abstract}

\maketitle

\section{ Introduction }

From the three prominent forms of relativistic Hamiltonian dynamics 
presented by Dirac~\cite{Dirac1949}, the point form (PF) is the 
least explored. However, it possesses virtues that are worth exploiting 
for the study of relativistic composite systems. One of the most 
important features is that of the 10 generators of the Poincar\'e 
algebra, those forming the Lorentz subgroup, rotations and boosts, are kinematic 
(free of interactions). This is to be contrasted to the most familiar 
instant-form (IF), where the boost operators make  changes of 
reference frames challenging, since they carry interaction terms 
(they are dynamical). This is particularly problematic in quantum 
field theories, where the number of particles is not conserved.~\footnote{
Ref.~\cite{Rocha:2009xq} shows how boosting
bound states in QCD using instant-form boosts becomes rather intractable.} 
The front form (FF), despite being the form with the larger 
kinematical group (containing 7 generators), has the drawback that 
 rotations are interaction dependent, and thus it makes the 
addition of angular momentum of relativistic interacting particles 
troublesome~\cite{Keister:1991sb,GomezRocha:2012hc}.  

In the last years, a considerable number of articles have been 
written with the goal of developing a new formalism able to describe 
the structure of hadrons --or, more generally, of relativistic
bound states-- in terms of the properties of their constituents by 
using the point form of relativistic quantum mechanics (PFRQM)
(cf.~\cite{Krassnigg:2003gh,Krassnigg:2004sp,
Biernat:2009my,Biernat:2010tp,Biernat:2014dea,GomezRocha2012,
Biernat:2011mp,Rocha:2010wm,Gomez-RochaThesis,GomezRocha:2012hca,
Gomez-Rocha:2013zma,Kleinhappel:2013tza,Kleinhappel:2012zj,Senekowitsch2014}). 
Relativistic quantum mechanics, unlike quantum field theory, considers 
a restricted number of degrees of freedom.  Poincar\'e invariance 
is ensured in this formalism by using the Bakamjian-Thomas 
construction~\cite{Bakamjian:1953kh,Keister:1991sb}. 
Its point-form version introduces a free velocity operator, that is 
multiplied by the interacting mass operator and leads to an 
interaction-dependent four-momentum operator~\cite{Keister:1991sb,Biernat:2010tp}.
Using a coupled-channel approach for that mass operator, we can 
describe the physical process from which invariant amplitudes and 
hadronic currents can be calculated.

An appropriate description of the structure of the current
poses several problems, and it is not straightforward to derive 
electroweak currents with all the required properties. Two 
basic features are Poincar\'e covariance and cluster separability~\cite{Sokolov:1977ym,
Coester:1982vt,Keister:1991sb,Siegert1937yt}. Our formalism in 
PFRQM has helped to understand the electroweak structure
of hadrons in several ways. It was initially applied to calculate 
the spectrum and decay widths of vector and axial-vector mesons within 
the chiral constituent quark model~\cite{Krassnigg:2003gh,Krassnigg:2004sp}. 
Later, electromagnetic properties of spin-0 and spin-1 two-body 
bound states with equal-constituent masses were studied~\cite{Biernat:2009my,
Biernat:2011mp,Biernat:2014dea}. More recently, the relativistic 
multichannel formalism was extended to unequal-mass constituents and 
to weak decay form factors in the time-like momentum transfer 
region~\cite{GomezRocha2012,GomezRocha:2012hca,Rocha:2010wm}. 
An additional condition that has to be satisfied by systems of 
unequal-constituent masses is to respect the heavy-quark symmetry 
predictions in the extreme case in which one of the masses is infinitely 
heavier than the other~\cite{Isgur1989ed,Isgur1989vq,NeubertHQS}. 
It was shown that our approach respects the required heavy-quark 
symmetry predictions~\cite{GomezRocha2012,Gomez-RochaThesis}, 
i.e. relations between electromagnetic and weak form factors appear 
in the limit $m_Q\to \infty$. This guarantees that the formalism 
is general enough to be applied to systems of arbitrary constituent masses, 
and gives us the freedom to apply it to heavy-to-heavy, heavy-to-light 
and light-to-light meson transitions. The main goal of this paper is 
to provide the result of the application of our formalism to all these cases.  
Ref.~\cite{GomezRocha2012} provides the basis and the main motivation for the 
present work. 

There is a second important issue we would like to address. It is known 
that the Bakamjian-Thomas construction produces problems related 
to cluster separability~\cite{Keister:1991sb}, which enter the 
calculation of form factors and may lead to unphysical contributions 
in the electromagnetic current~\cite{Biernat:2010tp,Biernat:2011mp,
Biernat:2014dea}. It was observed in~\cite{GomezRocha2012} that this 
is not the case in time-like processes, such as the weak decays we 
are going to consider in this work. Weak form factors can be extracted 
unambiguously and there is no need to introduce any additional spurious 
contribution to ensure the required covariant properties of
the hadronic currents. This does not mean, however, that the cluster 
problem is not present. The cluster problem is intrinsic to the 
Bakamjian-Thomas construction and we do not know any relativistic quantum 
mechanical approach that eliminates the cluster problem completely~\cite{Keister:2011ie}.

The need for additional covariants in space-like processes is similar to the occurrence 
within the covariant light-front formulation of Carbonell et al.~\cite{Carbonell:1998rj}, 
in which the orientation of the light front has to be considered explicitly in order to 
render the front-form approach manifestly covariant. In fact, comparisons between the 
point- and the front-form electromagnetic form factors show that the number of needed 
spurious contributions in spin-0 and spin-1 two-body systems coincides in the PF and FF 
cases~\cite{Biernat:2011mp,GomezRocha:2011qs}. 
In FF, one way to cure this problem is the introduction of 
pair-creation currents~\cite{Glazek:1989rr,deMelo:2005cy,deMelo:2010sw,Bakker:2003up,Simula:2002vm}, 
such as the so-called $Z$-graphs. 
This is particularly necessary in FF when one considers time-like processes, where
it is not possible to use the very convenient $q^+=0$. 
For $q^+>0$, additional covariants associated with zero modes 
are necessary in order to provide the appropriate Lorentz structure
of the weak current and a certain frame dependence of the form factors is 
encountered in processes that involve spin-1 mesons when the latter are considered as 
simple valence $q\bar q$ bound states~\cite{Bakker:2003up,Simula:2002vm,Jaus:1999zv,
Jaus:2002sv,Choi:2011xm,Choi:2010be,Cheng:1996if}. 
These problems are closely related 
to the violation of rotational invariance in the calculation
of one-body-current matrix elements in FF.

As mentioned above, in PF we do not encounter such kind of covariance problems in the current
for time-like momentum transfers and there is no need for introducing spurious contributions. Thus,
it is now interesting to consider a detailed numerical comparison between the point- and the front-form
results for time-like processes. For this comparison, we choose the light-front quark model of 
Ref.~\cite{Cheng:1996if}, and use the same harmonic-oscillator wave function and adopt the same 
harmonic-oscillator and mass parameters. Non-valence contributions are not considered explicitly 
in this work, neither in the work of Ref.~\cite{Cheng:1996if}. Since non-valence contributions 
enter differently in every form of dynamics, it is expected that considering the meson as a 
valence quark-antiquark pair only must result in different resulting form factors as well. 

The purpose of this paper is therefore twofold: on the one hand we apply the PFRQM approach to 
several particular cases of semileptonic decays using the harmonic-oscillator wave 
function that was used in previous works and obtain results that can be compared with 
experiments and with other approaches. With this we do not intend to make accurate predictions, but
rather to explore the applicability of our PFRQM approach to a broader range of reactions.
On the other hand, we perform a numerical comparison 
with an analogous front-form approach in which, as in our case, no additional nonvalence 
contributions are considered explicitly. Our purpose is to pose the question about the different 
role that non-valence contributions such as $Z$-graphs play in each role. 
The encountered differences reflect the fact that effects coming from 
vacuum fluctuations have to be treated differently in each form.

This article is organized as follows. Section II condenses the most important steps in the 
procedure used by the PFRQM approach and applies it to the process of a general weak semileptonic decay. 
In Section III we present and analyze our numerical results obtained in several particular cases 
for pseudoscalar mesons decaying into a pseudoscalar meson ($P\to P$) as well
as to vector mesons ($P\to V$). We compare our results with the analogous front-from approach
and discuss the encountered numerical differences. Conclusions and outlook are presented in Section IV.
Two important concepts of this formalism, \textit{velocity states} and \textit{vertex operators},  
are presented in Appendices A and B, respectively.

\section{Relativistic multichannel formalism and hadron currents}

The starting point of the derivation of currents and form factors in PFRQM 
is the physical processes in which such form factors are measured. In this 
work we examine semineptonic decays of mesons. In order to describe the 
processes in a fully Poincar\'e-invariant way, a multichannel version of 
the Bakamjian-Thomas construction~\cite{Bakamjian:1953kh,Keister:1991sb} 
is employed. In the point form version of the Bakamjian-Thomas construction 
the four-momentum operator factorizes into an interacting mass operator 
and a free four-velocity operator:
\begin{equation}\label{P=MV}
 \hat P^\mu = \hat P^\mu_{\text{free}} +
 \hat P^\mu_{\text{int}} = \hat M \hat V_{\text{free}}^\mu
 =(\hat M_{\text{free}}+\hat M_{\text{int}})\hat V_{\text{free}}^\mu.
\end{equation}
The four-velocity operator is free of interactions and is defined by 
$\hat V_{\text{free}}^\mu:=\hat P_{\text{free}}^\mu/\hat M_{\text{free}}
=\hat P^\mu/\hat M$. It describes the overall motion of the system. 
The mass operator $\hat M$, which depends on internal variables only,
is the quantity of interest, since it contains the information  of the 
internal structure of the system. 

The procedure to calculate invariant amplitudes of hadronic reactions, 
from which currents and form factors can be extracted, has been elaborately 
explained along several works, in which both electromagnetic and weak 
decays are considered (see~\cite{Biernat:2009my,GomezRocha2012,
Biernat:2014dea} for illustration, and~\cite{Biernat:2011mp,Gomez-RochaThesis} 
for deep details). We summarize here the most important steps that 
are required for the study of the processes in which we are interested,
and will refer to the more extended literature when necessary.

We will consider $P\to P$ meson transitions as well as 
$P\to V$ meson transitions.

\subsection{Derivation of the optical potential and identification of hadronic currents}

Our point-form approach is a coupled-channel formalism for a Bakamjian-Thomas 
mass operator formulated in the point form of Hamiltonian dynamics. Due to the 
form of Eq.~(\ref{P=MV}) the problem reduces to solve an eigenvalue equation 
for the mass operator $\hat M$: 
\begin{equation}\label{eigenvalue}
 \hat M|\psi\rangle =m |\psi\rangle,
\end{equation}
\noindent where $\hat M$ is the coupled channel mass operator for the Bakamjian-Thomas
construction in its point-form version. For a weak process of a meson $\alpha$ decaying 
into another meson $\alpha'$ the mass operator $\hat M$ needs -- at least -- four channels. 
They are needed in order to account for two possible time-ordered contributions, which 
are depicted in Fig.~\ref{DecayTimeOrderings}:
\begin{figure}[h!]
\begin{center}
\includegraphics[width=0.5\textwidth]{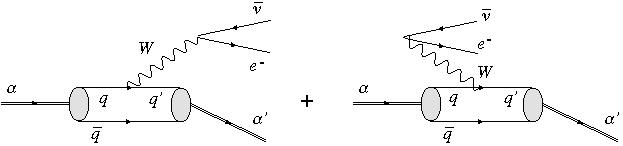}
\caption{Time-ordered contributions to the semileptonic decay of a 
meson $\alpha$ to $\alpha'$.}\label{DecayTimeOrderings}
\end{center}
\end{figure}
%
%
%
\begin{equation}\label{eq:massopdecay}
\hat M= 
 \left(\begin{array}{cccc} \hat M^{\mathrm{conf}}_{q\bar{q}} & 0 &
 \hat{K}_{q' \bar{q}W\rightarrow q\bar{q}} &
 \hat{K}_{q\bar{q}We\bar{\nu}_e\rightarrow q\bar{q}}\\
0 & \hat M^{\mathrm{conf}}_{q'\bar{q}e\bar{\nu}_e} &
\hat{K}_{q'\bar{q}W\rightarrow q' \bar{q}e\bar{\nu}_e} &
\hat{K}_{q\bar{q}We\bar{\nu}_e\rightarrow q' \bar{q}e\bar{\nu}_e} \\
\hat{K}^\dag_{q' \bar{q}W\rightarrow q\bar{q}} &
\hat{K}^\dag_{q' \bar{q}W\rightarrow q' \bar{q}e\bar{\nu}_e} & \hat
M^{\mathrm{conf}}_{q' \bar{q}W} &
0 \\
\hat{K}^\dag_{q\bar{q}We\bar{\nu}_e\rightarrow q\bar{q}} &
\hat{K}^\dag_{q\bar{q}We\bar{\nu}_e\rightarrow q' \bar{q}e\bar{\nu}_e}
& 0 & \hat M^{\mathrm{conf}}_{q\bar{q}We\bar{\nu}_e}
\end{array}\right)\, .
\end{equation}

The mass eigenstate $|\psi\rangle$ on which $\hat M$ acts
is a direct sum of $|\psi_{q\bar q} \rangle$, $|\psi_{q'\bar q e\bar \nu_e} \rangle$, 
$|\psi_{q'\bar q W} \rangle$ and $|\psi_{q\bar q W e\bar \nu_e} \rangle$ Hilbert spaces. 
In point form it is convenient to use a \textit{velocity-states} basis~\cite{Klink:1998zz,
Carbonell:1998rj,Karmanov:1998jp}, defined in Eq.~(\ref{Vstates}). The non-diagonal 
elements of $\hat M$ are \textit{vertex operators}, $\hat K^\dagger $ and $\hat K $, that 
describe the emission and absorption of the exchanged $W$-boson, respectively.
They are appropriately related to the weak interaction Lagrangian 
density through Eq.~(\ref{Vertex}) (see Refs.~\cite{Klink:2000pp,Biernat:2010tp} and 
App.~\ref{SecVertexLagrangian}). The instantaneous confining $q\bar q$ interaction is 
included in the diagonal elements of the matrix, which are denoted by ``conf'' 
(see~\cite{Biernat:2009my,GomezRocha2012}). For instance,
\begin{eqnarray}\label{eq:veigenst}
{M^{\mathrm{conf}}_{q\bar{q}W} \vert\,  \underline{v};
\vec{\underline{k}}_W, \underline{\mu}_W;
\vec{\underline{k}}_\alpha,\underline{\mu}_\alpha, \alpha \rangle}
= (\omega_{\underline{k}_W} +
\omega_{\underline{k}_\alpha} ) \vert\, \underline{v};
\vec{\underline{k}}_W, \underline{\mu}_W;
\vec{\underline{k}}_\alpha,\underline{\mu}_\alpha,  \alpha \rangle
\, , \\ && \nonumber
\end{eqnarray}
\noindent where $\underline{\mu}_\alpha$ denotes the spin orientation of the confined $q\bar q$ bound
state and $\alpha$ represents the remaining discrete quantum numbers necessary to specify it uniquely.
The energy of the $q\bar q$ bound state with quantum numbers $\alpha$ and mass $m_\alpha$ is represented by 
$\omega_{\underline{k}_\alpha}$ and expressed below Eq.~(\ref{eq:vcompl}). Underlined velocities, 
momenta and spin projections distinguish states with a confined $q\bar q$ pair from those with a free 
$q\bar q$ pair, which are not underlined.

The system of equations~(\ref{eigenvalue}) can be transformed into an equation for $|\psi_{q\bar q} \rangle$
by means of a Feshbach reduction, leading to the required expression for the optical potential that describes 
the entire process of the $W$-boson exchange, including both time-ordered contributions 
(cf. Fig.~\ref{DecayTimeOrderings}):
\begin{equation}
 \left(\hat M^{\mathrm{conf}}_{q\bar{q}} + 
 \hat V_{\mathrm{opt}}^{q\bar{q}\rightarrow q'\bar{q}e\bar{\nu}_e}(m) \right)
 |\psi_{q\bar q} \rangle=
 m |\psi_{q\bar q} \rangle,
\end{equation}
\noindent where
\begin{equation}
\hat V_{\mathrm{opt}}^{q\bar{q}\rightarrow q'\bar{q}e\bar{\nu}_e}(m)
= \hat{K}_{q'\bar{q}W\rightarrow q'\bar{q}e\bar{\nu}_e}
(m-M^{\mathrm{conf}}_{q' \bar{q}W})^{-1}\hat{K}^\dag_{q' \bar{q}W\rightarrow
q\bar{q}} + \hat{K}_{q\bar{q}We\bar{\nu}_e\rightarrow
q' \bar{q}e\bar{\nu}_e} (m-\hat
M^{\mathrm{conf}}_{q\bar{q}We\bar{\nu}_e})^{-1}
 \hat{K}^\dag_{q\bar{q}We\bar{\nu}_e\rightarrow q\bar{q}}\, .
\end{equation}

On-shell matrix elements of such optical potential have the structure of the 
invariant $\alpha \to \alpha'^{(*)} e \bar \nu_e$ decay amplitude resulting 
from leading-order covariant perturbation theory.  The calculation requires 
the insertion, in the appropriate places, of the spectral decomposition of 
the unity operators, written in the velocity-states basis (cf. App.~\ref{SecVelocityStates}). 
Since the calculation is tedious, it is not presented here in detail. We 
refer to Ref.~\cite{Gomez-RochaThesis} for technicalities, where the required 
matrix elements are given explicitly. From the structure of the invariant decay
amplitude it is straightforward to identify the microscopic hadron current:

\begin{eqnarray}\label{eq:voptcov}
&&{\langle \underline{v}^\prime; \vec{\underline{k}}_e^\prime,
\underline{\mu}_e^\prime; \vec{\underline{k}}_{\bar{\nu}_e}^\prime;
\vec{\underline{k}}_{\alpha^\prime}^\prime,\underline{\mu}_{\alpha^\prime}^\prime,
\alpha^\prime \vert \hat V_{\mathrm{opt}}^{b\bar{d}\rightarrow
c\bar{d}e\bar{\nu}_e}(m)\vert \vec{\underline{k}}_{\alpha},
\underline{\mu}_{\alpha}, \alpha \rangle_{\mathrm{os}}}
= \underline{v}_0 \delta^3 (\vec{\underline{v}}^{\, \prime} -
\vec{\underline{v}}\, )\, \frac{(2 \pi)^3
}{\sqrt{(\omega_{\underline{k}_e^{\prime}}+\omega_{\underline{k}_{\bar{\nu}_e}^{\prime}}
+\omega_{\underline{k}_{\alpha^\prime}^{\prime}})^3} \sqrt{
\omega_{\underline{k}_\alpha^{\phantom{\prime}}}^3}}\nonumber\\
&& \quad \times
\frac{e^2}{2\sin^2\vartheta_{\mathrm{w}}}V_{cb}\,
\frac{1}{2}\underbrace{\bar{u}_{\underline{\mu}_e^\prime}
(\vec{\underline{k}}_e^\prime)\gamma^\mu (1-\gamma^5)
v_{\underline{\mu}_{\bar{\nu}_e}^\prime}
(\vec{\underline{k}}_{\bar{\nu}_e}^\prime) }_{j_{\nu_e\rightarrow
e}^\mu(\vec{\underline{k}}_e^\prime,
\underline{\mu}_e^\prime;\vec{\underline{k}}_{\bar{\nu}_e}^\prime,
\underline{\mu}_{\bar{\nu}_e}^\prime)}
\frac{(-g_{\mu
\nu})}{(\underline{k}^\prime_e+\underline{k}^\prime_{\bar{\nu}_e})^2-m_W^2}
\frac{1}{2}{J_{\alpha\rightarrow
\alpha^\prime}^\nu(\vec{\underline{k}}_{\alpha^\prime}^\prime,
\underline{\mu}_{\alpha^\prime}^\prime;\vec{\underline{k}}_\alpha,
\underline{\mu}_\alpha)}\, .\nonumber
\end{eqnarray}

\noindent
where $\vartheta_{\mathrm{w}}$ is the electroweak mixing angle, $e$
the elementary electic charge and  $V_{cb}$ the  Cabibbo-Kobayashi-Maskawa
matrix element occurring at the $Wbc$-vertex. Note that the covariant structure 
of the $W$ propagator is only achieved when the sum of both time ordering 
contributions in Fig.~\ref{DecayTimeOrderings} are considered~\cite{Gomez-RochaThesis}. 

The semileptonic current extracted from Eq.~(\ref{eq:voptcov}) in the cases of a $P\to P$
transition has the structure:


\begin{eqnarray}\label{eq:Jwkpsps}
 J^{\nu}_{\alpha\rightarrow \alpha'}
(\vec{\underline{k}}_{\alpha'}^\prime;\vec{\underline{k}}_\alpha=\vec{0})&=&
\frac{\sqrt{\omega_{\underline{k}_\alpha}\omega_{\underline{k}_{\alpha'}^{\prime}}}}{4
\pi} \int\, \frac{d^3\tilde{k}_{\bar{q}}^\prime}{2 \omega_{k_q}}\,
\sqrt{\frac{\omega_{\tilde{k}^\prime_{q'}}+\omega_{\tilde{k}^\prime_{\bar{q}}}}
{\omega_{{k}^\prime_{q'}}+\omega_{{k}^\prime_{\bar{q}}}}} \,
\sqrt{\frac{\omega_{\tilde{k}_q} \omega_{\tilde{k}_{\bar{q}}}}
{\omega_{\tilde{k}^\prime_{q'}} \omega_{\tilde{k}^\prime_{\bar{q}}}}}
 \,  \bigg\{\!\sum_{\mu_q,\mu_{q'}^\prime
=\pm \frac{1}{2}}\!\!\!
\bar{u}_{\mu_{q'}^\prime}(\vec{k}_{q'}^\prime)\,\gamma^\nu\,(1-\gamma^5)\,
u_{\mu_q}(\vec{k}_q)\nonumber  \\ &&\times
D^{1/2}_{\mu_q\mu_{q'}^\prime}\!
\left[\!R_{\mathrm{W}}\!\left(\frac{\tilde{k}_{\bar
q}^\prime}{m_{\bar q}}, B_c(v_{{q'}\bar{q}}^\prime)\right)\,
\!R^{-1}_{\mathrm{W}}\!\left(\frac{\tilde{k}_{q'}^\prime}{m_{q'}},
B_c(v_{{q'}\bar{q}}^\prime) \right)\right] \bigg\}\,
\psi^\ast_{\alpha'}\,(\vert \vec{\tilde{k}}_{\bar{q}}^\prime\vert)\, \psi_\alpha
 \,(\vert \vec{\tilde{k}}_{\bar{q}}\vert)\, ,
\end{eqnarray}

\noindent and for $P\to V$ transitions \footnote{In the sequel, 
an asterisk is used to label a meson with total spin 1.}

\begin{eqnarray}\label{eq:Jwkpsv}
 J^{\nu}_{\alpha\rightarrow \alpha'^{\ast}}
(\vec{\underline{k}}_{\alpha'^\ast}^\prime,\underline{\mu}_{\alpha'^\ast}^\prime;\vec{\underline{k}}_\alpha=\vec{0})&=&
\!\!
\frac{\sqrt{\omega_{\underline{k}_\alpha}\omega_{\underline{k}_{\alpha'^\ast}^{\prime}}}}{4
\pi} \int\, \frac{d^3\tilde{k}_{\bar{q}}^\prime}{2 \omega_{k_q}}\,
\sqrt{\frac{\omega_{\tilde{k}^\prime_{q'}}+\omega_{\tilde{k}^\prime_{\bar{q}}}}
{\omega_{{k}^\prime_{q'}}+\omega_{{k}^\prime_{\bar{q}}}}} \,
\sqrt{\frac{\omega_{\tilde{k}_q} \omega_{\tilde{k}_{\bar{q}}}}
{\omega_{\tilde{k}^\prime_{q'}} \omega_{\tilde{k}^\prime_{\bar{q}}}}}
 \,
 \bigg\{\!\sum_{\mu_b,\mu_{q'}^\prime,\tilde\mu_{q'}^\prime,\tilde\mu_{\bar
 q}^\prime=\pm \frac{1}{2}}
 \!\!\!\!\!\!
\bar{u}_{\mu_{q'}^\prime}(\vec{k}_{q'}^\prime)\,\gamma^\nu\,
(1-\gamma^5) u_{\mu_q}(\vec{k}_q)  \nonumber \\
&&\hspace{-4cm}\times \sqrt{2} (-1)^{\frac{1}{2}-\mu_q}
C^{1\mu^\prime_{\!\alpha'^\ast}}_{\frac{1}{2}\tilde\mu_{q'}^\prime\frac{1}{2}
\tilde{\mu}_{\bar{q}}^\prime}\, D^{1/2}_{\tilde\mu_{q'}^\prime
\mu_{q'}^\prime}\!
\left[\!R^{-1}_{\mathrm{W}}\!\left(\frac{\tilde{k}_{q'}^\prime}{m_{q'}},
B_c(v_{q'\bar{q}}^\prime)\right)\right]\,D^{1/2}_{\tilde\mu_{\bar{q}}^\prime
-\mu_q}\!\left[ \!R^{-1}
_{\mathrm{W}}\!\left(\frac{\tilde{k}^\prime_{\bar q}}{m_{\bar q}},
B^{-1}_c(v_{{q'}\bar{q}}^\prime)\right)\right] \bigg\}\,
\psi^\ast_{\alpha'^{\ast}}\,(\vert \vec{\tilde{k}}_{\bar{q}}^\prime\vert)\, \psi_\alpha
 \,(\vert \vec{\tilde{k}}_{\bar{q}}\vert)\, .
\end{eqnarray}


The procedure presented here yields expressions for the 
hadronic currents that satisfy the required covariant 
properties, i.e. they transform as four-vectors. The 
proof requires to transform the velocity states to the 
physical momenta via a canonical boost $B_c(v)$~\cite{Biernat:2009my}.

In order to proceed to extract the form factors by using the obtained 
current matrix elements we need to specify the system kinematics. 
We make the most natural choice, in which a meson $\alpha$, initially 
at rest, decays into a meson $\alpha'$ moving in the $x$ direction
with momenta $\kappa_{\alpha'}$:

\begin{equation}\label{Kinematics}
\underline{k}_\alpha= \left( \begin{array}{c} m_\alpha\\ 0 \\0\\0
\end{array}\right) \quad\hbox{and}\quad
\underline{k}_{\alpha'}= 
\left( \begin{array}{c} 
\sqrt{m_{\alpha'}^2+\kappa_{\alpha'}^{2}}\\
\kappa_{\alpha'}
\\0\\0
\end{array}\right)
\end{equation}

with

\begin{equation}
\kappa_{\alpha'}^2=\frac{1}{4m_\alpha^2}
(m_\alpha^2+m_{\alpha'}^2-\underline{q}^2)^2-m_{\alpha'}^2\,.
\end{equation}
The modulus of the $\alpha'$ meson center-of-mass 
momentum $\kappa_{\alpha'}=|\underline{\vec{k}}_{\alpha'}|$ is thus
restricted by $0\leq \kappa_{\alpha'}^2 \leq
(m_\alpha^2-m_{\alpha'}^2)^2/(4m_\alpha^2)$. 
The allowed values of the 4-momentum
transfer squared are then

\begin{equation}
0\leq \underline{q}^2 \leq (m_\alpha-m_{\alpha'})^2 \, .
\end{equation}

\subsection{Form factors}
\label{FF}

Form factors are obtained by equating matrix elements of the obtained 
hadronic currents to its most general decomposition in terms of covariants 
and Lorentz-invariant functions. An appropriate decomposition of the 
$P\to P$ current can be written as~\cite{Wirbel:1985ji}:

\begin{eqnarray}\label{eq:Jpspsphys}
{J^\nu_{\alpha\rightarrow
\alpha'}(\vec{\underline{p}}_{\alpha'}^\prime;\vec{\underline{p}}_\alpha)}
=\left( (\underline{p}_\alpha+\underline{p}_{\alpha'}')^\nu -
\frac{m^2_\alpha-m^2_{\alpha'}}{\underline{q}^2 }\, \underline{q}^\nu
\right)F_1(\underline{q}^2)
+ \frac{m_\alpha^2-m_{\alpha'}^2}{\underline{q}^2}\, \underline{q}^\nu
F_0(\underline{q}^2)\, ,
\end{eqnarray}

\noindent where $\underline{q}=(\underline{p}_\alpha-\underline{p}_{\alpha'})$ 
is the time-like, 4-momentum transfer. And

\begin{eqnarray}\label{eq:Jpsvphys}
{J^\nu_{\alpha\rightarrow
{\alpha'}^\ast}(\vec{\underline{p}}_{{\alpha'}^\ast}^{\prime},\underline\sigma^\prime_{{\alpha'}^\ast};
\vec{\underline{p}}_\alpha)}
&=& \frac{2 i\epsilon^{\nu\mu\rho\sigma}}{m_\alpha+m_{{\alpha'}^*}}\,\epsilon^*_\mu(\vec{\underline{p}}_{{\alpha'}^\ast}^{\prime}, \underline\sigma^\prime_{{\alpha'}^\ast})\, \underline{p}'_{{\alpha'}^\ast\rho}\, \underline{p}_{\alpha\sigma} \, V(\underline{q}^2)
- (m_\alpha+m_{{\alpha'}^*})\, \epsilon^{*\nu}(\vec{\underline{p}}_{{\alpha'}^\ast}^{\prime}, \underline\sigma^\prime_{{\alpha'}^\ast})\, A_1(\underline{q}^2)
  \nonumber\\&& 
+ \frac{\epsilon^*(\vec{\underline{p}}_{{\alpha'}^\ast}^{\prime}, \underline\sigma^\prime_{{\alpha'}^\ast}) \cdot \underline{q}}{m_\alpha+m_{{\alpha'}^*}}\,(\underline{p}_\alpha+\underline{p}_{{\alpha'}^\ast}')^\nu\, A_2(\underline{q}^2)
+2 m_{{\alpha'}^*}\,\frac{\epsilon^*(\vec{\underline{p}}_{{\alpha'}^\ast}^{\prime}, \underline\sigma^\prime_{{\alpha'}^\ast}) \cdot \underline{q}}{\underline{q}^2}\, \underline{q}^\nu \, A_3(\underline{q}^2)
  \nonumber\\
  &&
- 2m_{{\alpha'}^*}\,\frac{\epsilon^*(\vec{\underline{p}}_{{\alpha'}^\ast}^{\prime}, \underline\sigma^\prime_{{\alpha'}^\ast}) \cdot \underline{q}}{\underline{q}^2}\, \underline{q}^\nu\, A_0(\underline{q}^2)\, ,
\end{eqnarray}

\noindent
in the $P \to V$ case~\cite{Wirbel:1985ji}. 
$\epsilon^*(\vec{\underline{p}}_{\alpha'^\ast}^{\prime}, \underline\sigma^\prime_{\alpha'^\ast})$
is the polarization 4-vector of the ${\alpha'}^\ast$ meson and $A_3(\underline{q}^2)$ the linear 
combination

\begin{equation}
A_3(\underline{q}^2)= \frac{m_\alpha+m_{\alpha'^\ast}}{2 m_{\alpha'^\ast}}\, A_1(\underline{q}^2) - 
\frac{m_\alpha-m_{\alpha'^\ast}}{2 m_{\alpha'^\ast}} A_2(\underline{q}^2)\, .
\end{equation}

With the kinematics adopted in 
Eq.~(\ref{Kinematics}) the 
polarization vectors read:

\begin{eqnarray}\label{eq:Dpol}
\epsilon(\underline{\vec{k}}^\prime_{{\alpha'}^\ast},\pm 1)
&=&\frac{1}{\sqrt{2}}(\mp \frac{\kappa_{{\alpha'}^\ast}}{m_{{\alpha'}^\ast}}, 
\mp\sqrt{1+(\frac{\kappa_{{\alpha'}^\ast}}{m_{{\alpha'}^\ast}})^2},-i,0)\, , \nonumber\\ 
\epsilon(\underline{\vec{k}}^\prime_{{\alpha'}^\ast},0 )&=&(0,0,0,1)\, .
\end{eqnarray}

The calculation  of the form factors requires the insertion of  
the expressions found in Eq.~(\ref{eq:Jwkpsps}) and in Eq.~(\ref{eq:Jwkpsv}), 
in the left-hand sides of Eqs.~(\ref{eq:Jpspsphys}) and~(\ref{eq:Jpsvphys}), 
respectively (see Ref.~\cite{GomezRocha2012} for details). Note that 
Eq.~(\ref{eq:Jpsvphys}) expresses a system of 4 equations with 4 unknowns 
for every polarization vector $\epsilon(\vec k, \mu)$, where $\mu$ can be 
1, -1 or 0. The kinematics used in Eq.~(\ref{Kinematics}) leads to 10 
non-vanishing matrix elements, namely $J^2(0)$, $J^3(0)$, $J^\mu(\pm 1)$, 
$\mu=0,1,2,3$, where $J^\nu(\underline\mu^\prime_{\alpha'^\ast}) := 
J^\nu_{\alpha\rightarrow \alpha'^\ast}(\vec{\underline{k}}_{\alpha'^\ast}^\prime,
\underline\mu^\prime_{\alpha'^\ast};\vec{\underline{k}}_\alpha)$.
Since $J^\mu(1)$ and $J^\mu(-1)$ are simply related by a space reflection one is 
left with 6 matrix elements of the current, from which only 4 
are independent. Consequently, the form factors are determined uniquely. 

This is an important achievement, since analogous calculations in the front 
form of dynamics fail in the attempt to extract the form factors 
unambiguously in $P\to V$ meson transitions. Difficulties originated by 
violation of rotational invariance in the front form make the description 
of the dynamics of spin-1 mesons troublesome. We will refer to this issue later.

It was demonstrated and explained in~\cite{GomezRocha2012}, that in time-like 
processes there is no manifestation of cluster separability violation that may 
appear in the form of non-physical contribution to the decomposition of the 
current. That occurred, by contrast, in the electromagnetic case~\cite{Biernat:2009my,
Biernat:2014dea}, and the problem was attributed to the cluster-separability 
violation caused by the Bakamjian-Thomas construction~\cite{Keister:1991sb}.
 
We are now in the position to present our numerical results for several weak 
decays and discuss the comparison with analogous results in the  front form 
of dynamics.

\section{Numerical studies}

The method presented here to derive hadronic currents and to extract form 
factors has been tested in~\cite{GomezRocha2012}. The work presented in 
Ref.~\cite{GomezRocha2012} extended the application of the PF formalism 
to the weak interaction, and considered mesons with different constituent-quark 
masses. Those studies allowed to see, in a comprehensive way, how the predictions 
of heavy-quark symmetry arise when one of the quark masses increases asymptotically.
As predicted by heavy-quark symmetry, electromagnetic and weak form factors are related 
in the exact limit $m_q\to \infty$. Such examinations provided analytic evidence 
for the expected connection between these two types of interactions, and consequently, 
a sign of reliability for our treatment of relativistic composite systems of different 
constituent masses. Ref.~\cite{GomezRocha2012} focused on the study of heavy-quark 
symmetry as well as on cluster-separability properties of this point-form approach. 
For that purpose the same harmonic-oscillator wave function with parameter 
$a=0.55$ GeV was used for all numerical studies.

In the present work, however,  we want to test the PFRQM approach in another way. 
We introduce a flavor dependence in the wave function, by assuming a different 
harmonic-oscillator parameter for each meson. Even taking into account this flavor 
dependence, the model remains very simple. Thus, it might not be sophisticated enough
to establish quantitative predictions which could be compared with experiments. 
Nonetheless, it is necessary to carry out such calculations for several decays in 
order to understand how the point-form approach compares with other approaches and 
to learn at least qualitatively how the transition form factors depend on the kind 
of transition considered. Numerical studies within such a simple model will serve as 
our starting point for future developments in PFRQM. 

We are particularly interested in comparisons with front-form results and in the 
role of non-valence contributions in the description of currents and form factors. 
In front form, such non-valence contributions turn out to become important when one 
goes from space-like to time-like momentum transfers, and thus they may play a role 
in the point-form approach as well. For time-like momentum transfer it is not possible 
to use the $q^+=0$ frame in front form. As a consequence, non-valence configurations 
leading to $Z$-graph contributions (quark-antiquark pairs created from the vacuum) can 
occur. Such $Z$-graph contributions have been analyzed in Ref.~\cite{Bakker:2003up}. 
Applying analytic continuation ($q_\perp \to i q_\perp$) from the space-like to the 
time-lime momentum transfer region to the transition form factors calculated in a 
$q^+=0$ frame for space-like to the time-like region (where $q^+\neq 0$), provided 
that the $Z$-graph contributions are appropriately taken into account. The importance 
of the $Z$-graph contributions decreases with increasing the mass of the heavy quark 
and it vanishes in the heavy-quark limit, since an infinitely heavy quark-antiquark 
pair cannot be produced out of the vacuum~\cite{Simula:2002vm}. The numerical values 
obtained for the Isgur-Wise function within the point form approach agree with those 
obtained within the analogous front-form quark model~ \cite{GomezRocha2012}. As soon 
as the decay form factors are calculated for finite physical masses of the heavy quarks, 
differences between the point- and front-form approach must appear.

Another --but related-- particular issue we would like to address in the context of 
these comparisons concerns the frame dependence that appears in the calculation of 
form factors of $P\to V$ transitions in the front-form approach. In the light-front 
quark model of Ref.~\cite{Cheng:1996if}, the authors choose a frame in which the 
momentum transfer is purely longitudinal, i.e. $q_\perp=0$, $q^2=q^+q^-$. Working 
in this way, form factors of processes that involve vector mesons cannot be extracted 
unambiguously, and the form factors exhibit a dependence on whether the daughter 
meson goes in the positive or negative $z$-direction. On the other hand, it was shown 
in Ref.~\cite{GomezRocha2012} that in the point form there is no frame dependence of 
the form factors in time-like processes and they can be determined unambiguously form 
the different components of the current. 

In order to quantify all these differences, let us define first the wave function and 
the parameters employed in these numerical studies.

 \subsection{Meson wave function}

The form factors are solely determined by the $q\bar q$ bound-state wave function
and the constituent quark masses. One is free to use any model wave function obtained 
from a particular bound-state problem. We choose the harmonic-oscillator wave function
defined as:

\begin{equation}\label{eq:wavefunc}
\psi(\kappa)=
\frac{2}{\pi^{\frac{1}{4}}a^{\frac{3}{2}}}
\exp\left(-\frac{\kappa^2}{2a^2}\right)\,,
\end{equation}

\noindent 
which allows us for a  direct comparison with Ref.~\cite{Cheng:1996if}. The numerical 
results presented here have been computed using the model parameters quoted in 
Table~\ref{ParametersFlavorDependent}, which have been taken from Ref.~\cite{Cheng:1996if} 
as well.

\begin{table}[h]
\begin{center}
\begin{tabular}{cccccccc cccc}
   $a_\pi$  &  $a_\rho$  & $a_K$  & $a_{K^*}$  & $a_D$ & $a_{D^*}$  & $a_B$  & $a_{B^*}$ & 
   $m_{u,d}$ & $m_b$   & $m_c$ & $m_s$\\     
   \hline 
   \hline
     \,\, 0.33\,\,  & \,\,0.30\,\, & \,\,0.38\,\, & \,\,0.31\,\,  & \,\,0.46\,\,  & \,\,0.47\,\,  & \,\,0.55\,\,  & \,\,0.55\,\, &
     \,\,0.25\,\,   & \,\,4.8\,\,  & \,\,1.6\,\,  & \,\,0.40\,\,  \\ \hline
\end{tabular}
\caption{Harmonic-oscillator parameters and quark masses (in GeV) used for the calculation 
of transition form factors in this work. They were determined in Ref.~\cite{Cheng:1996if} 
by fitting the wave functions to the experimental values for the decay constants. The 
Cabibbo-Kobayashi-Maskawa matrix element $|V_{cb}|$ as well as the physical meson masses 
are those quoted by the Particle Data Group~\cite{Beringer:1900zz}.}\label{ParametersFlavorDependent}
\end{center}
\end{table}

\subsection{$P\to P$ transitions}  

For pseudoscalar-to-pseudoscalar transitions, 
in order to allow for comparison with other 
works, besides $F_0(q^2)$  and $F_1(q^2)$, also 
$f_-(q^2)$ is depicted for all computed decays. 
$f_-(q^2)$ and $f_+(q^2)$ are defined by 
\begin{equation}
 J^\mu(p_1,p_2) = f_+(q^2) (p_1+p_2)^\mu +f_-(q^2) (p_1-p_2)^\mu,
\end{equation}
\noindent
where $p_1$ and $p_2$ are the initial and final meson 4-momenta. 
Their relation with $F_0(q^2)$  and $F_1(q^2)$ is given by:
\begin{equation}
 F_1(q^2)=f_+(q^2), \quad F_0(q^2)=f_+(q^2)+\frac{q^2}{{m_\alpha}^2-{m_{\alpha'}}^2}f_-(q^2).
\end{equation} 
The values at $q^2=0$ for $F_1(0)$, or equivalently for $f_+(0)$, 
are shown in Table~\ref{fplus} together with the results obtained 
within the light-front quark model~\cite{Cheng:1996if}. For 
heavy-to-heavy transitions, i.e. $B\to D$, as well as for 
$B\to\pi$ transitions, both FF and PF
results seem to agree quite well, whereas they differ slightly 
for $D\to\pi(K)$. 

We do not have a definitive explanation for this fact,
but we suspect that these differences are due to the 
different way in which $Z$-graphs and other non-valence
contributions enter the form factors in either approach. 
There is a particular frame, namely the $q^+=0$ frame, 
in the front form, where $Z$-graphs disappear. In point 
form a particular $q^+=0$ frame can be realized for 
lepton-hadron scattering by taking the limit of infinitely 
large Mandelstam $s$, which corresponds to the infinite-momentum 
frame of the hadron (cf.~\cite{Senekowitsch2014,Gomez-Rocha:2013zma}). 
This explains, e.g., the equality of our 
point-form results for electromagnetic meson form factors 
(for $q^2<0$) with corresponding front-form results~\cite{Biernat:2009my,
GomezRocha2012}. In the $q^+=0$ frame however, weak 
decays cannot take place, since the process is necessarily 
time-like ($q^2=q^+q^- -q_\perp > 0$) or light-like at the 
point for maximal recoil ($q^2=0$). In the light-front quark 
model of Ref.~\cite{Cheng:1996if}, the calculations are done 
in a frame where the momentum transfer is purely longitudinal, 
this is $q_\perp=0$, $q^2=q^+q^-$.  At $q^2=0$ either $q^+$ or 
$q^-$ must vanish which corresponds to the daughter meson 
going either in + or in $-$ $z$-direction, respectively.
Since the pseudoscalar decay form factors do not depend on 
whether the daughter meson goes into + or $-$ $z$-direction, 
one can assume $q^+=0$. This implies, however, that 
$Z$-contributions vanish at the maximum recoil point. 
For $q^2>0$ there is, however, no argument to exclude $Z$-graph
contributions in the decay form factors. In point form one does 
not even have an argument at $q^+=0$ (apart of the mass of 
the produced $Q\bar Q$-pair) that $Z$-graphs should vanish. 

A quantitative estimate of the $Z$-graph contribution
is not within the scope of this work. We have seen, however, 
in the previous work of Ref.~\cite{GomezRocha2012} that the 
point-form results reproduce the front-form ones exactly in the 
heavy-quark limit. 
One can therefore
expect that for heavy-to-heavy transitions point-form and front-form 
results show a greater resemblance than for heavy-to-light transitions.
For heavy-to-light processes non-valence contributions are expected to 
be more important. It is thus not surprising that the results differ 
in both approaches.  In the $D\to K$ and $D\to \pi$ cases, point- and 
front-form results differ considerably, the front-form results being 
somewhat closer to the experimental data~\cite{Beringer:1900zz}. 

\begin{figure}[H]
\begin{center}
  \includegraphics[width=0.5\textwidth]{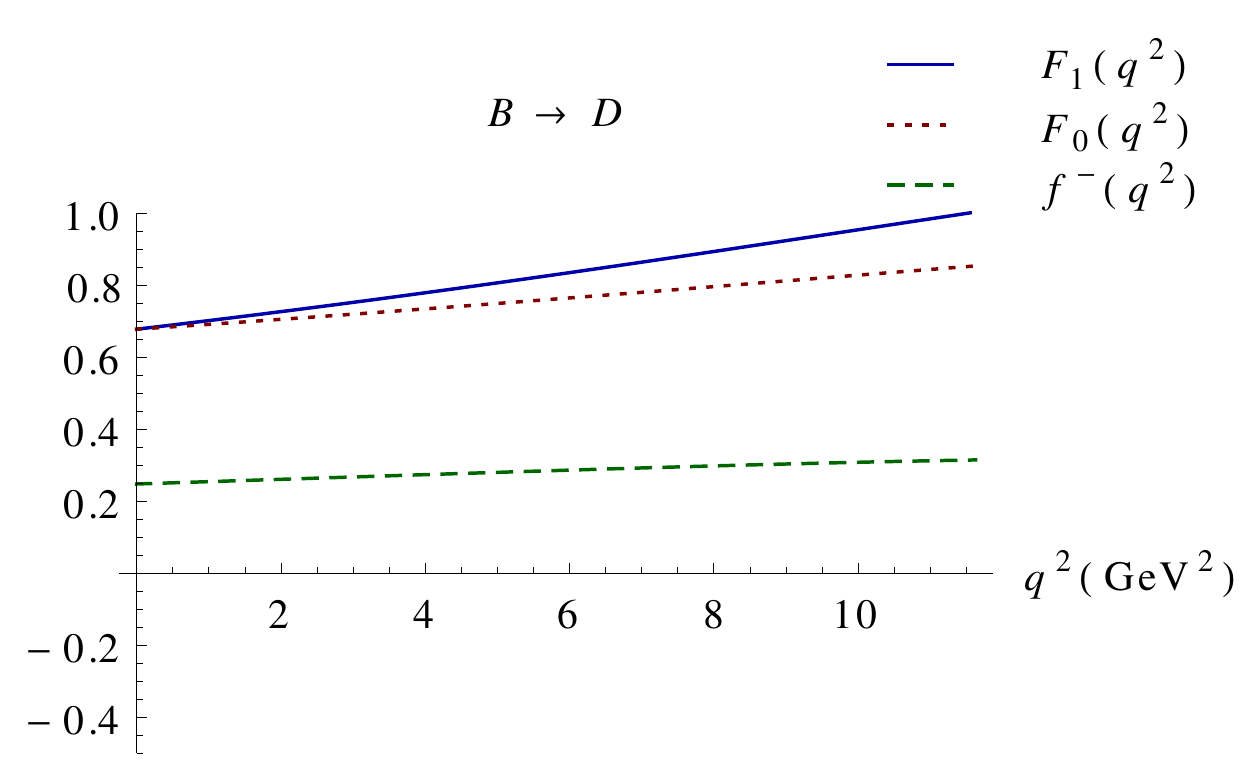}
  \includegraphics[width=0.49\textwidth]{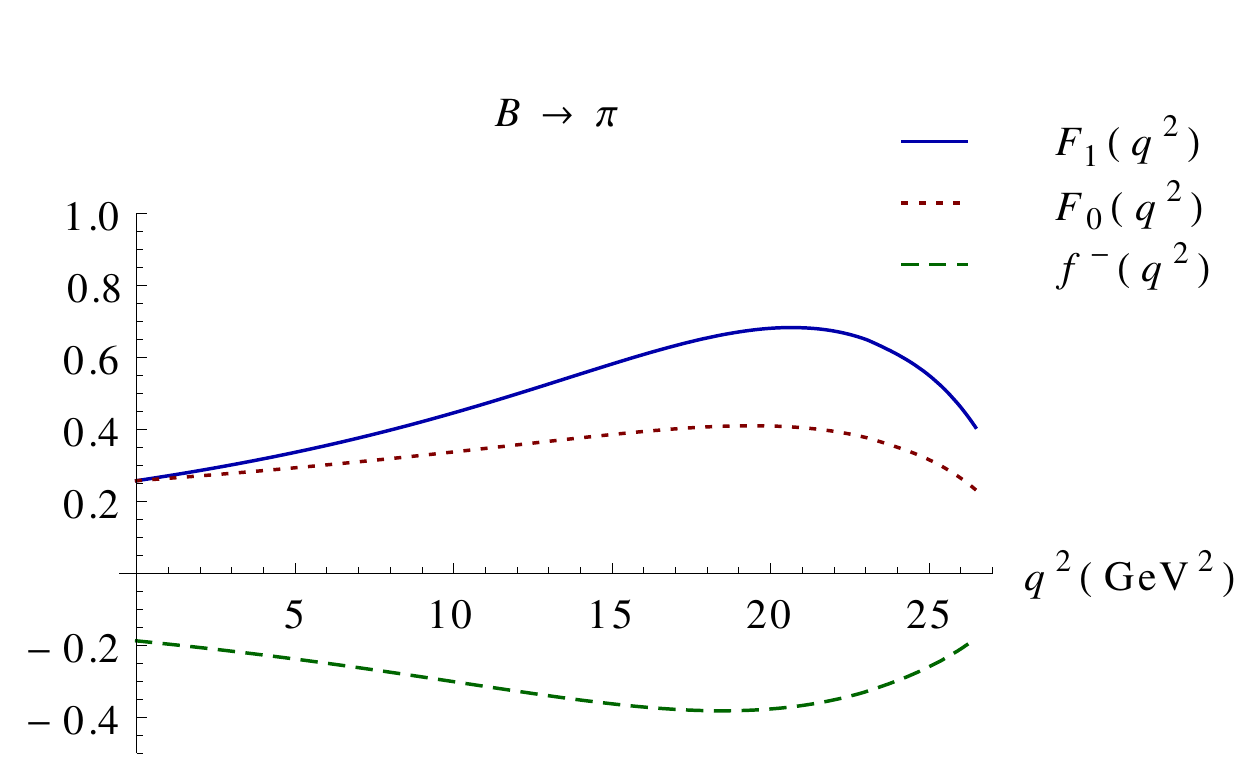}
\caption{$B\to D$ and $B\to \pi$ transition form factors in the whole range $0\leq q^2\leq(M_B-M_{D(\pi)})^2$. 
Parameters for the quark masses and harmonic-oscillator
wave functions are taken from 
Table~\ref{ParametersFlavorDependent}. 
For the meson masses the current values given by the Particle Data Group have been  
taken~\cite{Beringer:1900zz}.}\label{fig:BtoD}
\end{center}
  \end{figure}

\begin{figure}[H]
\begin{center}
  \includegraphics[width=0.49\textwidth]{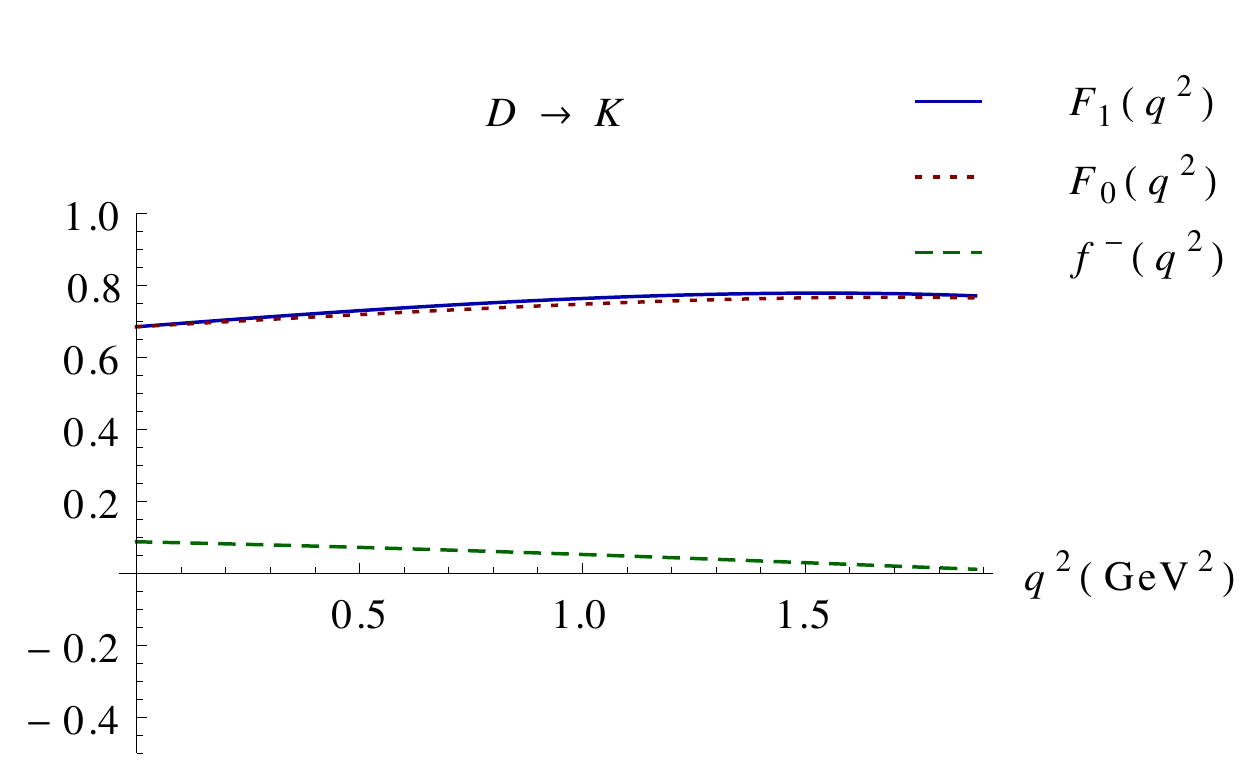}
  \includegraphics[width=0.49\textwidth]{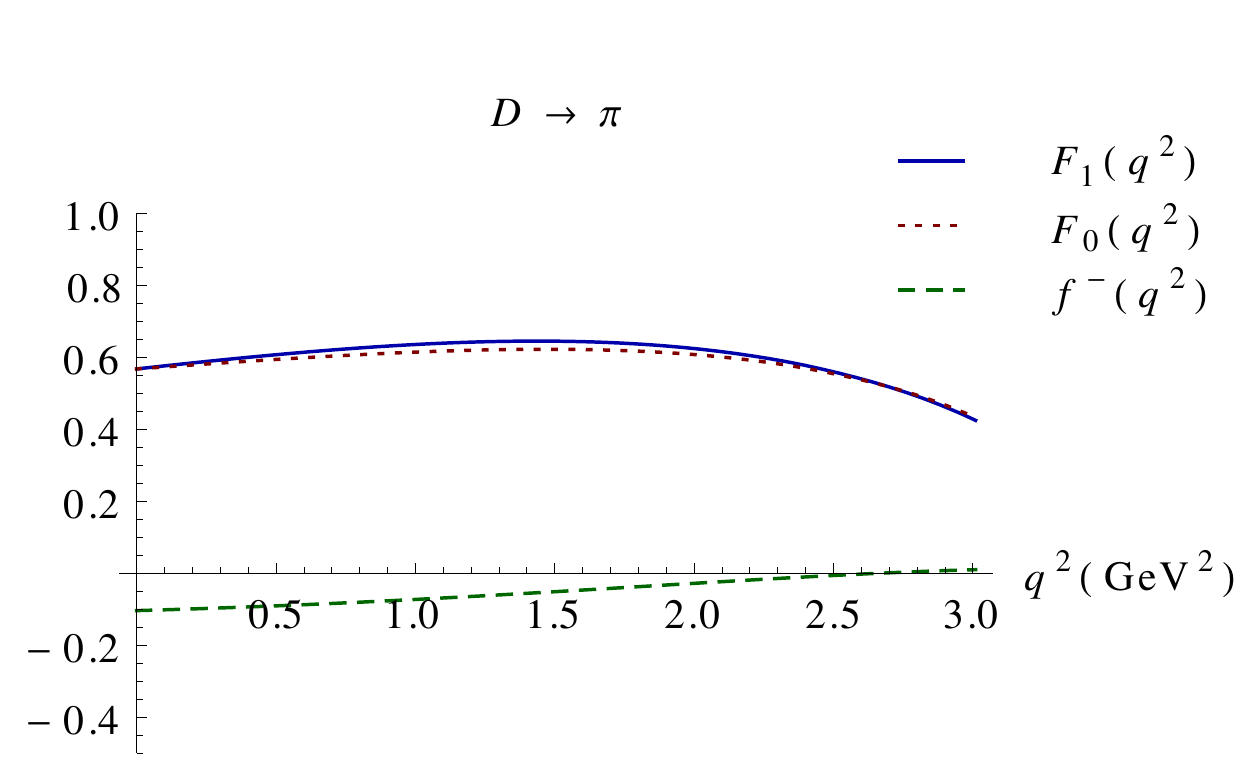}
\caption{
Same as in Fig.~\ref{fig:BtoD} for 
$D\to K$ and $D\to \pi$ transition form factors.}\label{fig:DtoPiK}
\end{center}
  \end{figure}

Another resemblance with the front-form results is that 
$f_-(q^2)\sim - f_+(q^2)$ for $B\to \pi$  and to less extent for 
$D\to \pi$ (cf. Figs.~\ref{fig:BtoD} and~\ref{fig:DtoPiK}).
Near zero recoil (where $q^2$ is maximal) heavy-quark symmetry predicts 
$(f_+ +f_-)^{B(D)\pi}\sim \frac{1}{\sqrt{m_{B(D)}}}$. In our case we have
\begin{equation}
(f_+ +f_-)^{B\pi}_{q_{\text{max}}}\sim 0.22,
\qquad (f_+ +f_-)^{D\pi}_{q_{\text{max}}}\sim 0.43,
\end{equation}
whereas $1/\sqrt{m_B}\sim 0.43$ and $1/\sqrt{m_D}\sim 0.73$.

\begin{table}[h]
\begin{center}
    \begin{tabular}{c|c|c|c}
    
	Decay   &    FF~\cite{Cheng:1996if} &  PF [this work]  & Exp.~\cite{Beringer:1900zz}  \\          \hline  \hline 
        $B\to D$   &  0.70      &  0.68  &  --  \\
	$B\to\pi$  &  0.26      &  0.26	 &   --  \\
	$D\to\pi$  &  0.64      &  0.57  & 0.661$\pm$0.022 \\
	$D\to K$   &  0.75      &  0.68  & 0.727$\pm$0.011 \\
	 \hline
    \end{tabular}\caption{$F_1(0)$, or equivalently $f_+(0)$, form factor for $P\to P$
transitions, corresponding to Figs.~\ref{fig:BtoD} - \ref{fig:DtoPiK}.}\label{fplus}
\end{center}
\end{table}
   
In order to get an idea of the reliability of the results 
as a function of the quality of the wave function, we have 
recomputed the above given results using a unique parameter 
$a=0.42$ GeV, which is the average of the values considered in this work. 
The form factors $F_1$ at $q^2=0$ are given in 
Table~\ref{F1aa042}. One can appreciate a considerable difference with 
respect to those given in Table~\ref{fplus}, as a consequence of the need 
of distinguishing the meson considered.

\begin{table}[h]
\begin{center}
    \begin{tabular}{ c c c c}
    	     \;$B\to D$\;  & \; $B\to\pi$\;  & \;$D\to\pi$ \;  & \;$D\to K$ \;  \\      \hline  \hline 
	     0.37       &   0.32      &   0.64      &  0.71      \\	
	\hline
    \end{tabular}\caption{ $F_1(0)$, or equivalently $f_+(0)$, form factor obtained for  $P\to P$
transitions, using the same oscillator parameter 
$a=0.42$ GeV.}\label{F1aa042}
\end{center}
\end{table}

\subsection{$P\to V$ transitions}\label{SecPtoV}

The comparison for transitions that involve mesons with spin 
is more interesting. In the light-front quark model~\cite{Cheng:1996if}, 
the form factors for $P\to V$ meson transitions  extracted in the $q_\perp=0$ frame,
exhibit a certain frame dependence. For a given $q^2$, the form factors depend on
whether the recoiling daughter moves in the positive ``+'' or negative ``$-$'' 
$z$-direction relative to the parent meson. In the light-front quark model the 
results for the form factors are larger in the ``+'' frame than in the ``$-$'' 
one.  The exact vanishing of $Z$-graphs at $q^2=0$ in the ``+'' frame is taken 
as an argument in Ref.~\cite{Cheng:1996if} to conclude that $Z$-graphs are less 
important in the ``+'' frame than in the ``$-$'' frame.

In Table~\ref{TableBtoDstar} results for both frames together with the point-form
results obtained in this work are given at $q^2=0$. The authors of~\cite{Cheng:1996if} 
interpret the difference between the results at $q^2=0$ in the ``+'' and ``$-$'' frames 
as a measure for the $Z$-graph contribution present in the ``$-$'' frame. In the point 
form all time-like form factors can be extracted without ambiguity and no frame dependence 
appears in our description of weak decays. Again, the scope of this work does not allow 
to give a precise estimate of $Z$-graph contributions. One could perhaps guess that they 
are of the same order of magnitude as the difference between ``+'' and ``$-$'' frames 
in front from.

In Tabs.~\ref{TableBtoDstar}-\ref{TableDtoRho} our from-factor results at $q^2=0$
are compared with those of Ref.~\cite{Cheng:1996if} for several decays. One observes 
that the results obtained in the point form for $A_0(0)$, $A_1(0)$ and $A_2(0)$ are 
very similar in all the computed transitions, whereas they differ notably in the front 
form. There seems to be a good agreement between both approaches for $V(0)$ and $A_0(0)$. 
For these two form factors one sees that for the heavy-to-heavy transition the point-form 
result lies between the obtained ones in the front form in the ``+'' and ``$-$'' frames, 
being closer to the ``+'' one. $A_1(0)$ and $A_2(0)$ turn out to be larger in the point
form in all cases. 

\begin{figure}[H]
\begin{center}  
 \includegraphics[width=0.485\textwidth]{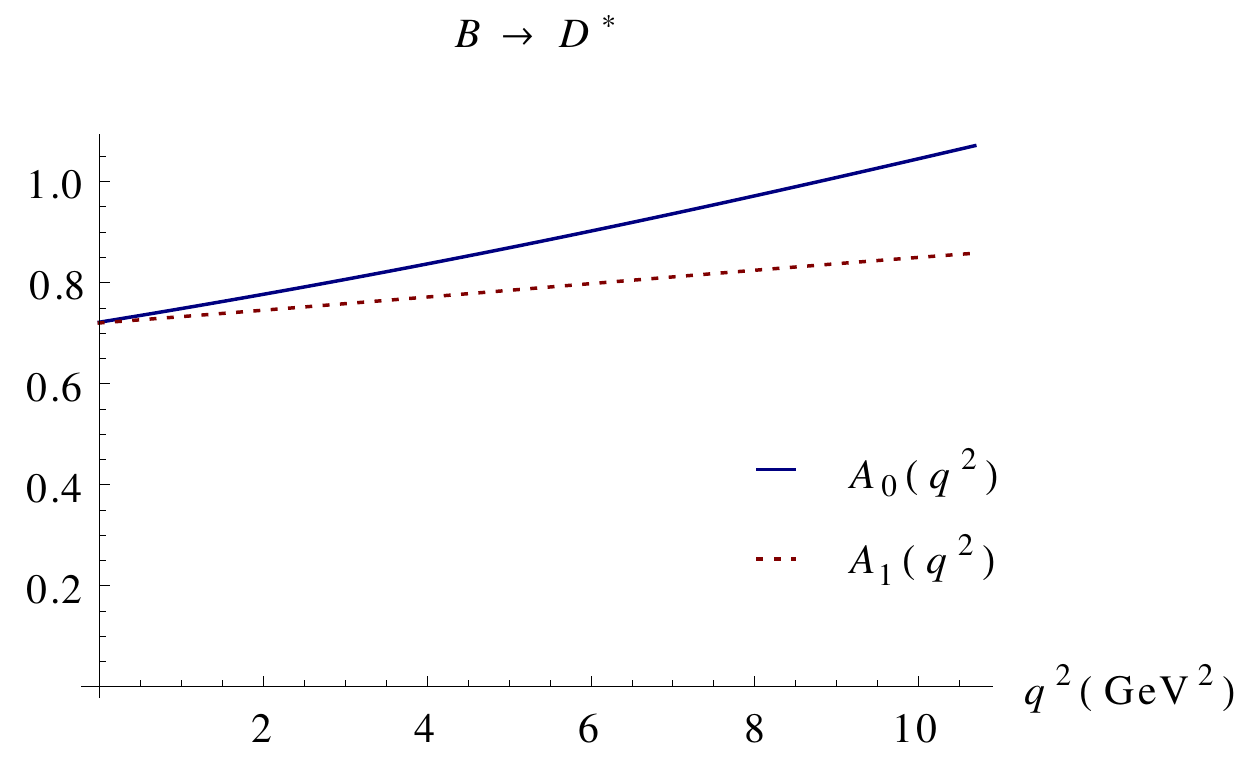}
  \includegraphics[width=0.485\textwidth]{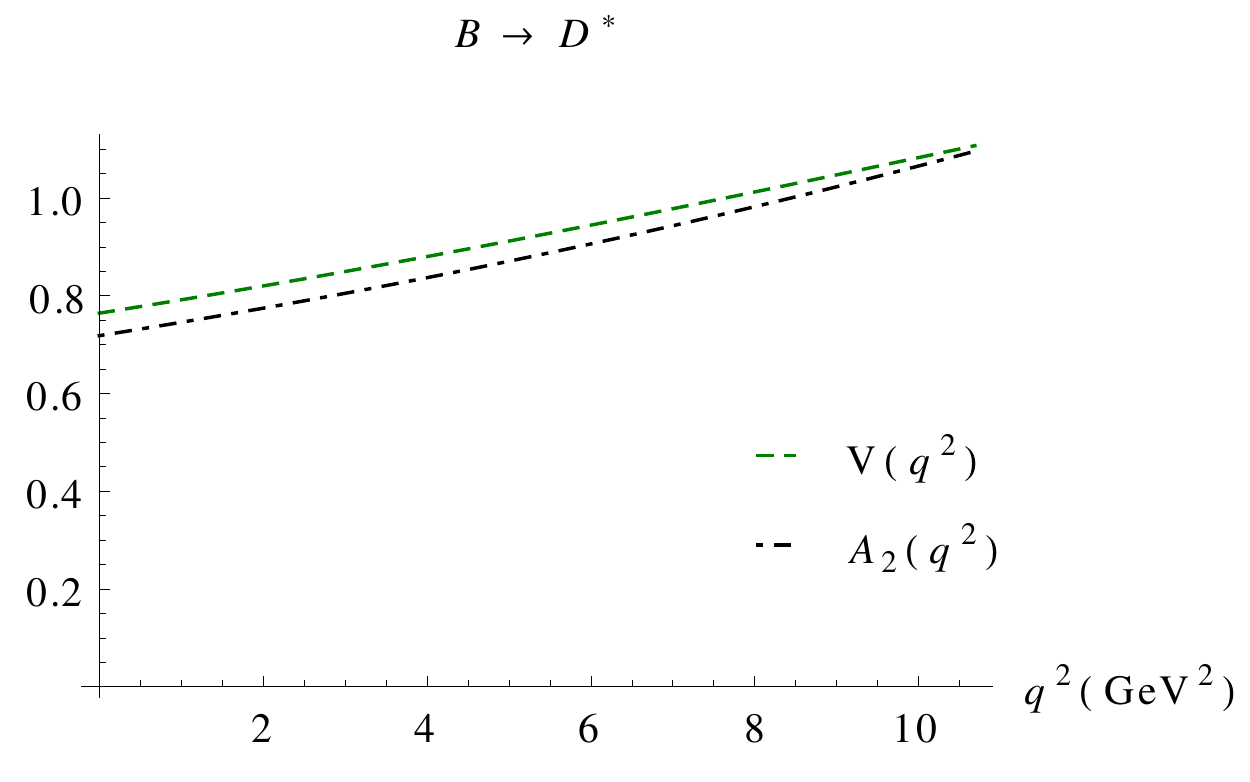}
\caption{$B\to D^*$ transition form factors in the whole range
 $0\leq q^2\leq(M_{B}-M_{D^*})^2$. 
Parameters for the quark masses and harmonic-oscillator
wave functions are taken from 
Table~\ref{ParametersFlavorDependent}. 
For the meson masses the current values given by the Particle Data Group are 
taken~\cite{Beringer:1900zz}.}\label{FigBtoD}
\end{center}
  \end{figure}

\begin{figure}[H]
\begin{center}
  \includegraphics[width=0.49\textwidth]{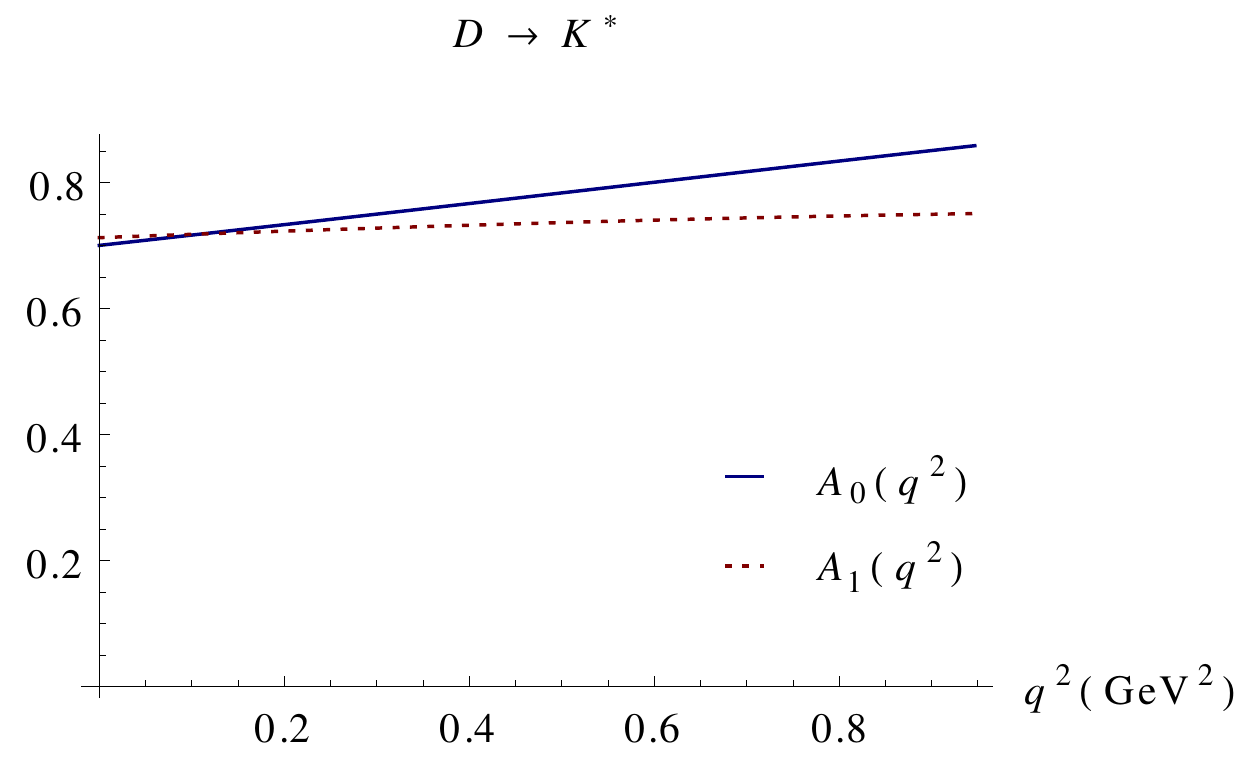}
  \includegraphics[width=0.49\textwidth]{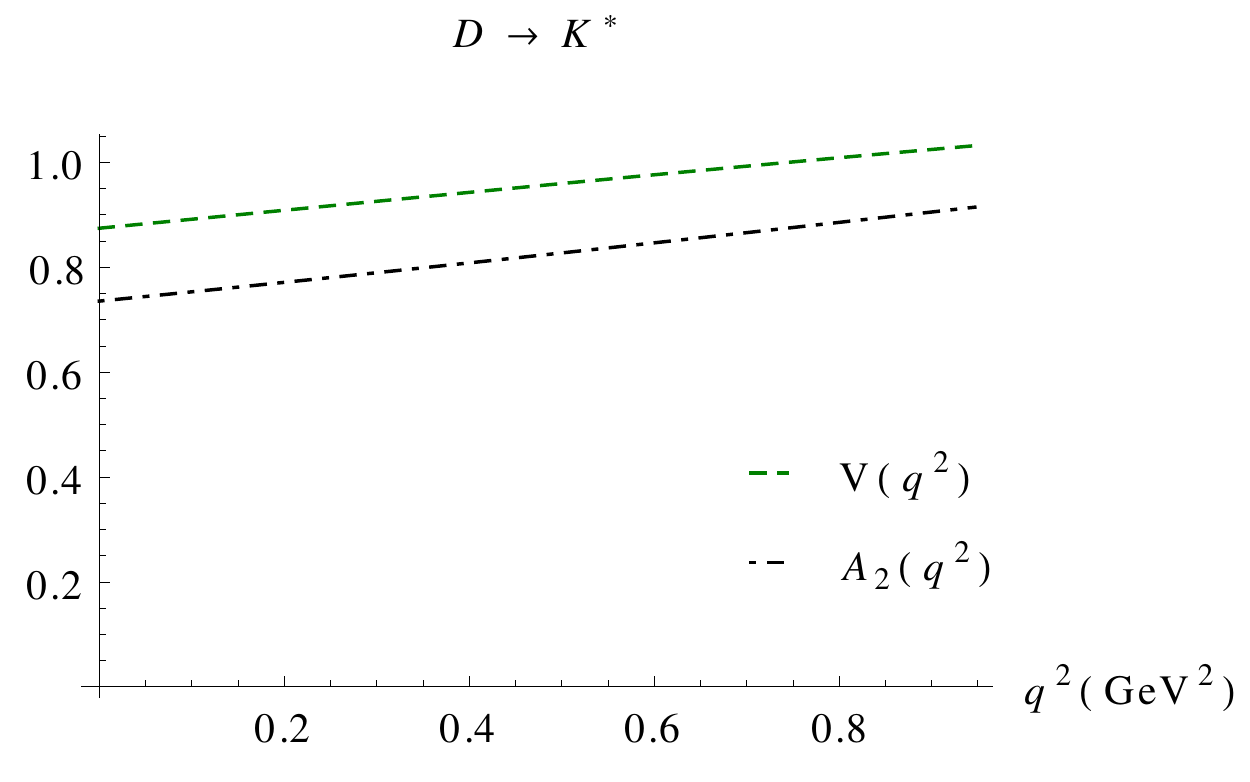}
\caption{
Same as in Fig.~\ref{FigBtoD} for 
$D\to K^*$ transition form factors.}\label{FigDtoKstar}
\end{center}
  \end{figure}

\begin{figure}[H]
\begin{center}
  \includegraphics[width=0.49\textwidth]{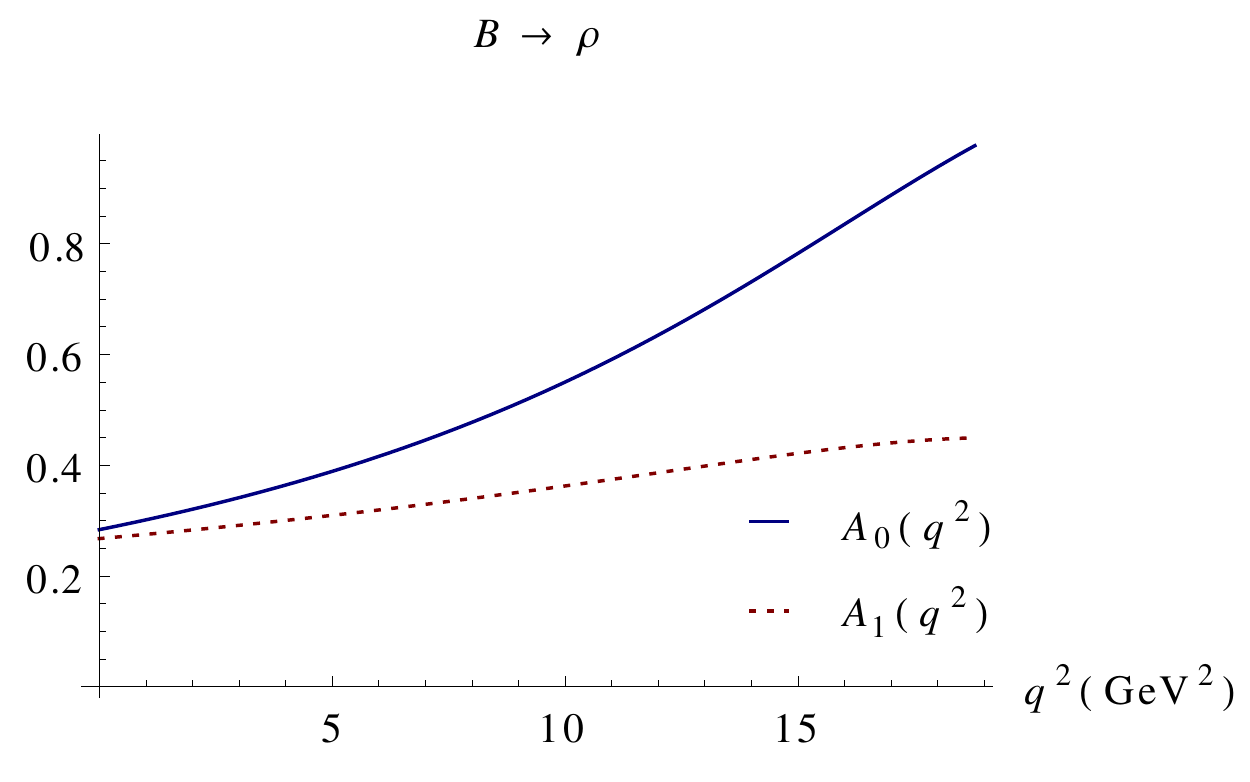}
  \includegraphics[width=0.49\textwidth]{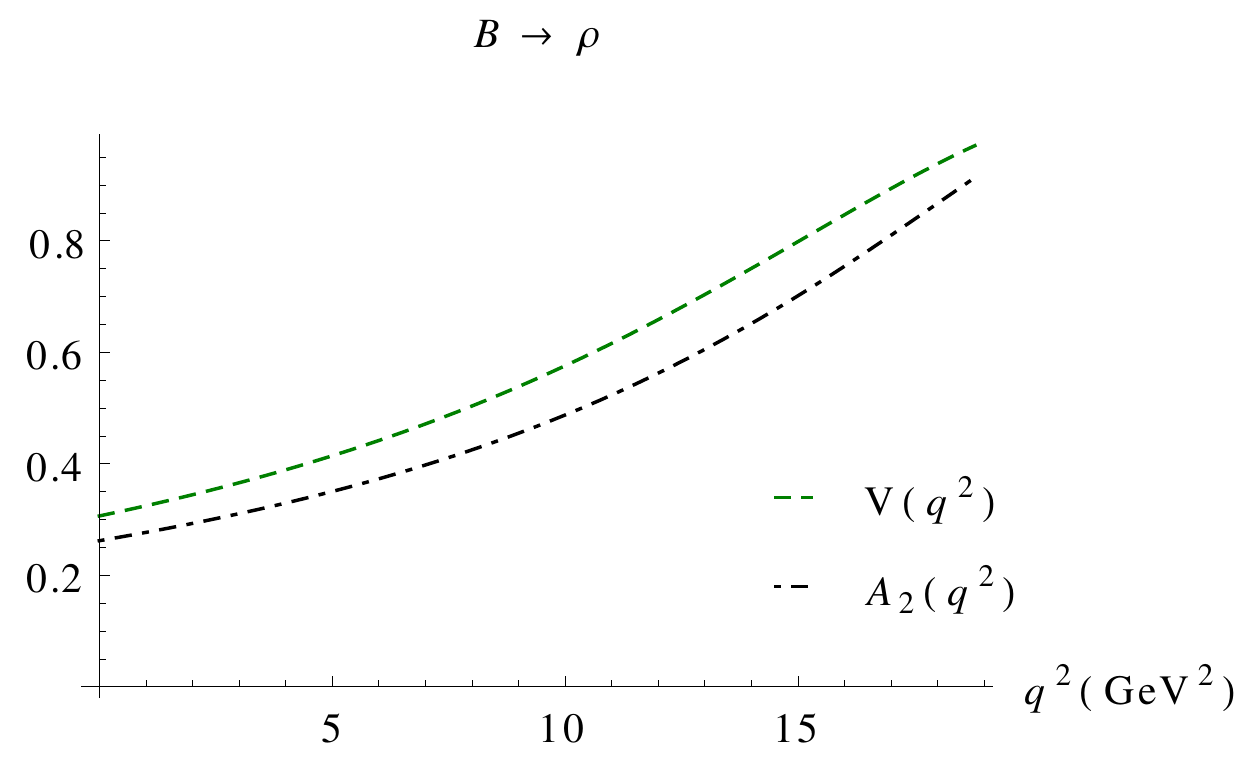}
\caption{
Same as in Fig.~\ref{FigBtoD} for
$B\to \rho$ transition form factors.}\label{FigBtorho}
\end{center}
  \end{figure}

\begin{figure}[H]
\begin{center}
  \includegraphics[width=0.49\textwidth]{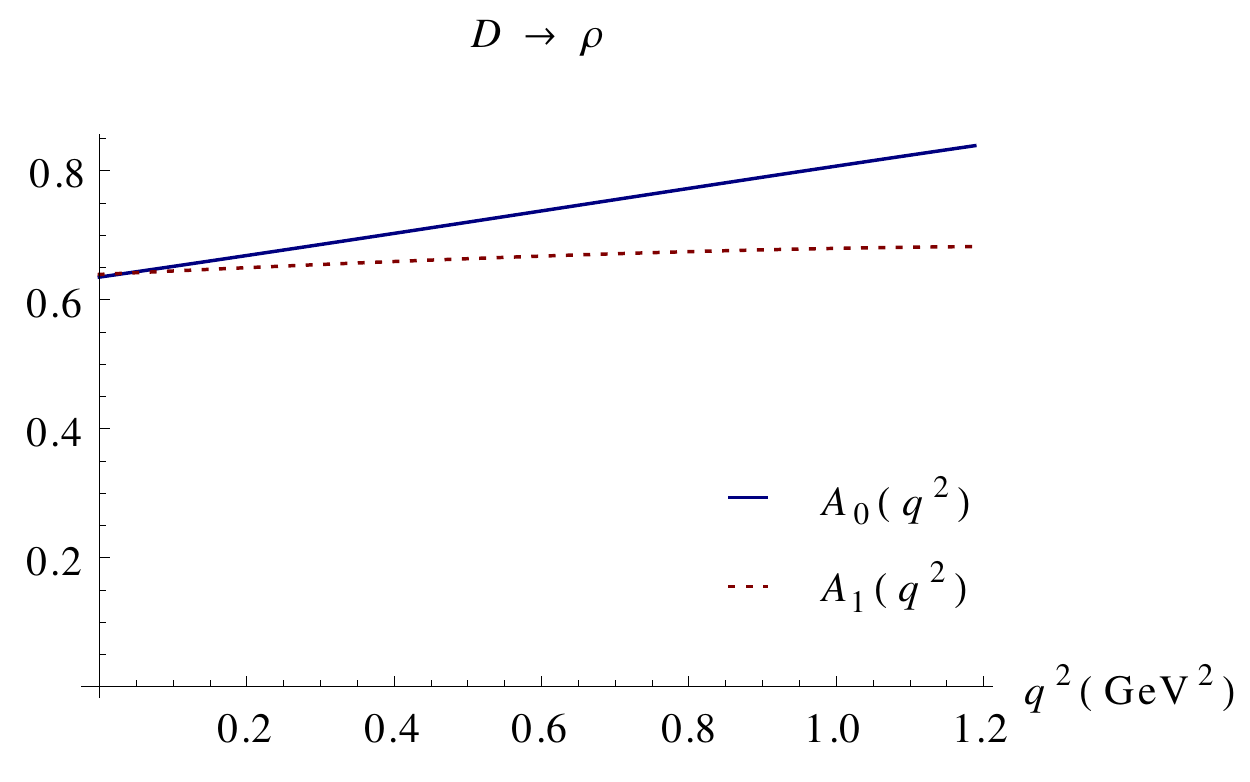}
  \includegraphics[width=0.49\textwidth]{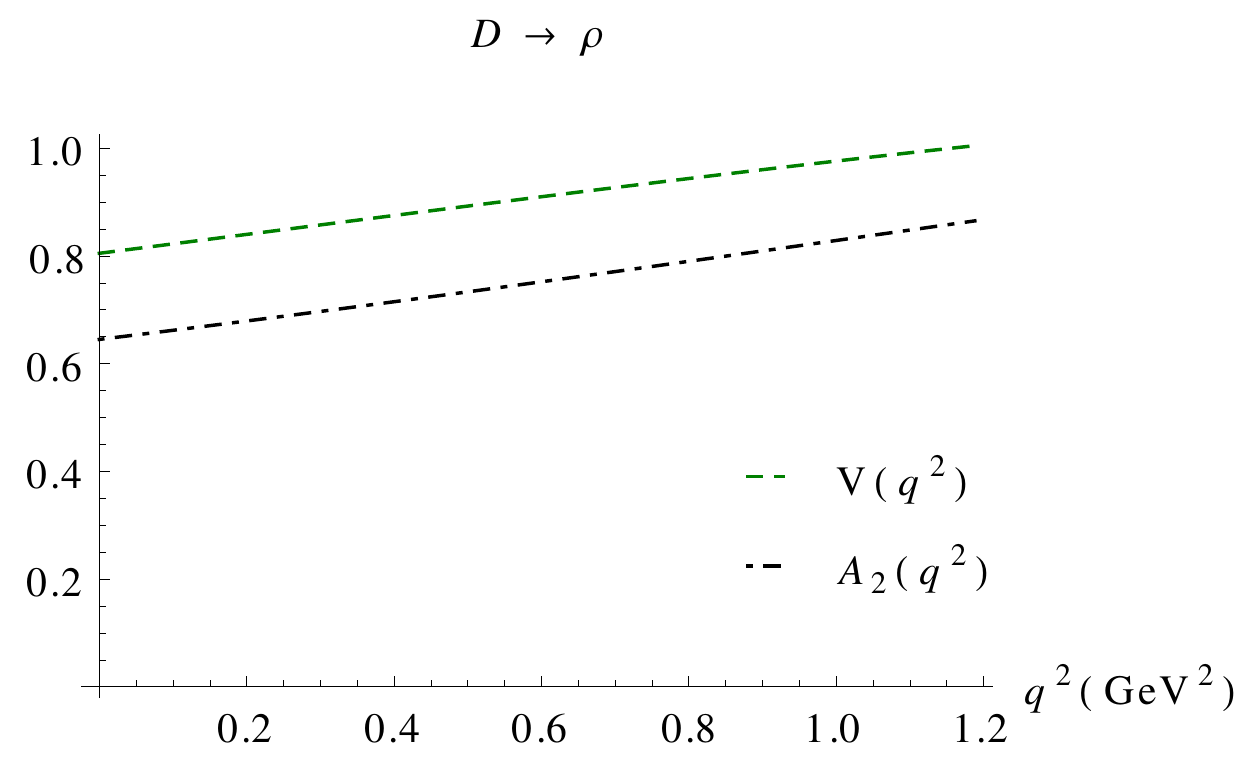}
\caption{
Same as in Fig.~\ref{FigBtoD} for
$D\to \rho$ transition form factors.}\label{FigDtorho}
\end{center}
  \end{figure}

For the whole $q^2$ range, i.e. $0\leq q^2\leq ({m_\alpha}-{m_{\alpha'}})^2$, the
form factors $V(q^2)$, $A_{0}(q^2)$, $A_1(q^2)$ and $A_2(q^2)$  are depicted in 
Figs.~\ref{FigBtoD}-\ref{FigDtorho}. If one compares with the corresponding plots in 
Ref.~\cite{Cheng:1996if} the observations made already for $q^2=0$ are confirmed. 
For the $B$ decays our form factors resemble very much those of Ref.~\cite{Cheng:1996if} 
(in the ``+'' frame) with $A_2(q^2)$ showing the biggest deviations. For $D$-decays larger 
differences can be observed, in particular for $A_1(q^2)$ and $A_2(q^2)$, but the qualitative 
features of the form factors are still quite similar. This discrepancy is, of course, 
foreseeable since the point- and front-form approaches are not equivalent as long as one does
not include non-valence contributions. The equivalence is only reached in the heavy-quark limit, 
where the same Isgur-Wise function is obtained~\cite{GomezRocha2012}.

\begin{table}
\begin{center}
    \begin{tabular}{c|cccc}
	 $B\to D^*$  &   $V(0)$ & $A_0(0)$ & $A_1(0)$ & $A_2(0)$    \\            \hline \hline
	FF~\cite{Cheng:1996if} in the ``+'' frame &    0.78     & 0.73 & 0.68 &   0.61  	   \\
FF~\cite{Cheng:1996if} in the ``$-$'' frame &  0.62 & 0.58 & 0.59 & 0.61\\
	PF (this work)  &     0.76   & 0.72 & 0.72 & 0.72           \\
	 \hline
    \end{tabular}\caption{Form factors at $q^2=0$ for the $B\to D^*$ transition obtained 
within the light-front quark model in Ref.~\cite{Cheng:1996if} (FF) in the frames where the recoiling daughter
moves in the positive  $z$-direction (``+'' frame) and negative  $z$-direction
(``$-$'' frame)
in comparison with the results obtained in the point form (PF).}\label{TableBtoDstar}
\end{center}

\begin{center}
    \begin{tabular}{c|cccc}
	 $D\to K^*$  &   $V(0)$ & $A_0(0)$ & $A_1(0)$ & $A_2(0)$    \\            \hline \hline
	FF~\cite{Cheng:1996if}     & 0.87 &  0.71     & 0.62  &  0.46 	   \\
	PF [this work]            & 0.87   &  0.70     & 0.71  &  0.73     \\
	 \hline
    \end{tabular}\caption{Form factors at $q^2=0$ for the $D\to K^*$ transition obtained 
within the light-form quark model (FF) in the frame where  where the recoiling daughter
moves in the positive  $z$-direction , i.e. ``+'' frame, and in the point form (FF)
of relativistic quantum mechanics.}\label{TableDtoKstar}
\end{center}

\begin{center}
    \begin{tabular}{c|cccc}
	 $B\to \rho$  &   $V(0)$ & $A_0(0)$ & $A_1(0)$ & $A_2(0)$    \\            \hline \hline
	FF~\cite{Cheng:1996if}     & 0.30  &  0.28     & 0.20  &  0.18 \\
	PF [this work]            & 0.31   &  0.28     & 0.27  &  0.26     \\
	 \hline
    \end{tabular}\caption{Same comparison as in Table~\ref{TableDtoKstar} but for 
the $B\to\rho$ transition.}\label{TableBtoRho}
\end{center}

\begin{center}
    \begin{tabular}{c|cccc}
	 $D\to \rho$  &   $V(0)$ & $A_0(0)$ & $A_1(0)$ & $A_2(0)$    \\            \hline \hline
	FF~\cite{Cheng:1996if}  &  0.78     & 0.63  &  0.51   & 0.34  	   \\
	PF [this work]          &  0.80     & 0.63  &  0.64   & 0.64       \\
	 \hline
    \end{tabular}\caption{Same comparison as in Table~\ref{TableDtoKstar} but for 
the $D\to\rho$ transition.}\label{TableDtoRho}
\end{center}
\end{table}

 
\section{Conclusions and outlook}

We have applied the PFRQM approach to several  weak decays.
Numerical results have been given, that can be compared with 
experiments and with other approaches, e.g.  the analogous 
calculation in front form considered herein. While the harmonic-oscillator 
wave function, Eq.~(\ref{eq:wavefunc}), still might be too simple to 
make quantitative predictions that can be compared with experiments,
it has served as a first step in the understanding of our point-form 
approach by means of numerical studies that allow for direct comparisons.

While in the front form the obtained results for $P\to V$ transitions 
exhibit a certain dependence on the reference frame, i.e. on whether the recoiling 
daughter moves in the positive or negative $z$-direction relative to the parent
meson, in the point form all form factors are determined unambiguously. 

Furthermore, in contrast to what happened using the approach in the 
electromagnetic case~\cite{Biernat:2011mp,Biernat:2014dea},  we are able to extract 
all form factors of mesons with spin 0 decaying into mesons of spin 0 and 
to spin 1  without the need of introducing any non-physical contribution 
--spurious form factor-- to correct covariant deficiencies of the current.

In the heavy-quark limit, as was shown in Ref.~\cite{GomezRocha2012}, 
point-form and front-form calculations yield the same numerical result 
for the Isgur-Wise function. This equivalence is possible because in the 
heavy-quark limit nonvalence contributions, such as $Z$-graphs vanish. 
Numerical comparisons of our outcome with analogous front form calculations 
show that the results obtained from both approaches do not coincide exactly
outside the limit. This is not surprising. Non-valence 
contributions  such as  $Z$-graphs cannot be present in the $m_Q\to \infty$ limit. 
On the other hand, such $Z$-graph contributions as well as other vacuum induced 
currents do not exist in the front form, even for finite masses if one choose
the $q^+=0$ frame, since momentum conservation imposes the ``+'' sum of momenta 
at every vertex to be positive. Such kind of non-valence contributions cannot 
be excluded, however, in the point form. As long as non-valence contributions 
are not calculated explicitly, the point- and the front-form approaches cannot 
be equivalent. 
 
From the coincidence of both approaches in the heavy-quark limit we 
conclude that the appeared discrepancy between the point and the front forms
outside of the heavy-quark limit must be due to the different way in which 
vacuum-induced currents enter the description in every form of dynamics.

All this is relevant in order to explore the effect of introducing 
additional degrees of freedom in the approach. It is the subject of future work
to introduce $Z$-graphs contributions explicitly in the coupled channel approach
and to investigate how they affect the form factors (for recent advances in this 
direction, see~\cite{Gomez-Rocha:2013zma}). Similar studies on this subject were 
carried out in the front form~\cite{Bakker:2003up}. Like in Ref.~\cite{Bakker:2003up} 
an estimate on $Z$-graph contributions within our approach could be obtained by 
calculating the transition form factors in the space-like region, where one can go 
into the infinite-momentum frame and continue those results analytically to the 
time-like momentum-transfer region. The work presented here poses the starting point 
for this goal. Studies concerning weak decays in the space-like region and its 
analytic continuation have been initiated recently in~\cite{Gomez-RochaThesis,Senekowitsch2014}.

\begin{appendix}

\section{Velocity states}
\label{SecVelocityStates}

An $n$-particle velocity state $|v;\vec k_1,\mu_1;\vec k_2,\mu_2;...;\vec k_n,\mu_n\rangle$
is defined through an overall velocity $v$ and $n$ individual momenta and spin 
projections $\{\vec k_i,\mu_i\}$, such that $\sum_{i=1}^n \vec k_i=0$. 
A velocity state represents an
$n$-particle system in the rest frame that is boosted to a frame with
a total 4-velocity $v$ ($v^\mu v_\mu=1$) by means of a canonical boost 
$B_c(v)$ \cite{Keister:1991sb}:

 \begin{eqnarray}\label{Vstates}
&&\vert v; \vec{k}_1, \mu_1; \vec{k}_2, \mu_2;\dots;
\vec{k}_n, \mu_n \rangle 
:= \hat{U}_{B_c(v)} \, \vert
\vec{k}_1, \mu_1; \vec{k}_2, \mu_2;\dots; \vec{k}_n, \mu_n \rangle.
\end{eqnarray}

They satisfy the orthogonality and completeness relations~\cite{Krassnigg:2003gh}:
\begin{eqnarray}\label{eq:vnorm}
&&\langle v^\prime; \vec{k}_1^\prime,
\mu_1^\prime; \vec{k}_2^\prime, \mu_2^\prime;\dots;
\vec{k}_n^\prime, \mu_n^\prime \vert \, v; \vec{k}_1, \mu_1;
\vec{k}_2, \mu_2;\dots; \vec{k}_n, \mu_n \rangle 
 \nonumber\\
 &&\quad
= v_0  \delta^3(\vec{v}^\prime-\vec{v}) \frac{(2 \pi)^3 2
\omega_{k_n}}{\left( \sum_{i=1}^n \omega_{k_i}\right)^3}
\left( \prod_{i=1}^{n-1} (2\pi)^3
2 \omega_{k_i} \delta^3(\vec{k}_i^\prime-\vec{k}_i)\right) 
\left(
\prod_{i=1}^{n} \delta_{\mu_i^\prime \mu_i}\right)\,\nonumber\\
\end{eqnarray}
and 
\begin{eqnarray}\label{eq:vcompl}
\mathbf{1}_{1,\dots,n}&=&\sum_{\mu_1=-j_1}^{j_1}
\dots \sum_{\mu_n=-j_n}^{j_n} \int
\frac{d^3 v}{(2\pi)^3 v_0}  \left[
\prod_{i=1}^{n-1}\frac{d^3k_i}{(2 \pi)^3 2 \omega_{k_i}}
\right]\nonumber\\
&&\times\frac{\left(\sum_{i=1}^n \omega_{k_i}\right)^3}{2
\omega_{k_n}} \vert v; \vec{k}_1, \mu_1;
\vec{k}_2, \mu_2;\dots; \vec{k}_n, \mu_n \rangle
 \langle v; \vec{k}_1, \mu_1; \vec{k}_2,
\mu_2;\dots; \vec{k}_n, \mu_n\vert\, ,\nonumber\\
\end{eqnarray}
with $m_i$, $\omega_{k_i}:= \sqrt{m_i^2+\vec{k}_i^2}$, and $j_i$,
being the mass, the energy, and the spin of the $i$th particle,
respectively. 

Velocity states transform under Lorentz transformations $\Lambda$ as

\begin{eqnarray}\label{eq:vstateboost}
\lefteqn{\hat{U}_\Lambda \vert v; \vec{k}_1, \mu_1; \vec{k}_2,
\mu_2;\dots; \vec{k}_n, \mu_n \rangle} \nonumber\\ &&=
\sum_{\mu_1^\prime, \mu_2^\prime,\dots,\mu_n^\prime}\, \left\{
\prod_{i=1}^n \, D^{j_i}_{\mu_i^\prime \mu_i}\left[R_{\mathrm
W}(v,\Lambda)\right] \right\}
 \vert \Lambda v; \overrightarrow{R_{\mathrm
W}(v,\Lambda)k}_1, \mu_1^\prime; \overrightarrow{R_{\mathrm
W}(v,\Lambda)k}_2, \mu_2^\prime;\dots; \overrightarrow{R_{\mathrm
W}(v,\Lambda)k}_n, \mu_n^\prime \rangle\, ,\nonumber\\
\end{eqnarray}
with the Wigner-rotation matrix
\begin{equation}\label{eq:wignerrot}
R_{\mathrm W}(v,\Lambda) = B_c^{-1}(\Lambda v)\Lambda B_c(v)\, ,
\end{equation}
\noindent where
\begin{equation}\label{canonical boost}
B_c(v) = \left( \begin{array}{cc} v^0 & \mathbf v^T \\ \mathbf v &
\mathbf 1 +\frac{v^0 -1}{\mathbf v ^2 }\mathbf v \mathbf v ^T
\end{array}\right).
\end{equation}

\section{Vertex Operators}
\label{SecVertexLagrangian}

The creation and annihilation of particles is introduced in this framework by means of 
\textit{vertex operators} $\hat K$ that are specified by the velocity-state representation
and an appropriate relation to the pertinent field-theoretical interaction-Lagrangian 
density $\hat{\mathcal{L}}_{\text{int}}$. In this work, $\hat{\mathcal{L}}_{\text{int}}$ corresponds to the 
Lagrangian density of the weak interaction. Due to velocity conservation that 
follows from the point-form version of the  Bakamjian-Thomas construction, one is led to 
define matrix elements of $\hat K$ by~\cite{Biernat:2010tp,Klink:2000pp}:
\begin{eqnarray}\label{Vertex}
&& \langle v,\vec k_1,\mu_1;...;\vec k_{n+1},\mu_{n+1}| 
\hat K^\dagger| v,\vec k_1,\mu_1;\vec k_2,\mu_2;...;\vec k_n,\mu_n\rangle
= \langle v,\vec k_1,\mu_1;\vec k_2,\mu_2;...;\vec k_n,\mu_n| 
\hat K|  v,\vec k_1,\mu_1;...;\vec k_{n+1},\mu_{n+1}\rangle ^*\nonumber\\
&&\quad=\mathcal{N}_{n+1,n}v^0\delta^3(\vec v-\vec v')
\langle v,\vec k_1,\mu_1;...;\vec k_{n+1},\mu_{n+1}| 
\hat{\mathcal L}_\text{int}(0)f(\Delta m)| v,\vec k_1,\mu_1;\vec k_2,\mu_2;...;\vec k_n,\mu_n\rangle,\nonumber\\
\end{eqnarray}
where $\mathcal{N}_{n+1,n}=(2\pi)^3/\sqrt{\mathcal M'_{n+1}\mathcal M'_{n}}$, 
$\mathcal M'_{n}=\sum_{i=1}^k \omega_{i}$ 
and $f(\Delta m=\mathcal M'_{n+1}-\mathcal M'_{n})$ denotes a vertex form factor 
that can be introduced in order to account for (part of) the neglected off-diagonal 
velocity contributions and to regulate integrals.

\end{appendix}

\begin{acknowledgments}
I would like to thank Wolfgang Schweiger 
and Oliver Senekowitsch for valuable discussions. 
I also thank Andreas Krassnigg for a critical reading 
of the manuscript. This work was supported by the 
Austrian Science Fund (FWF) via doctoral program  
DK W1203-N1 and project nr. P25121-N27.
\end{acknowledgments}



\begin{thebibliography}{18}



\bibliographystyle{acm}



\bibitem{Dirac1949} 
  P.~A.~M.~Dirac,
  Rev.\ Mod.\ Phys.\  {\bf 21}, 392 (1949).
 
 
\bibitem{Rocha:2009xq} 
  M.~Gomez~Rocha, F.~J.~Llanes-Estrada, D.~Schutte and S.~V.~Chavez,
  Eur.\ J.\ Phys.\ A {\bf 44}, 411 (2010)
  [arXiv:0910.1448 [hep-ph]].

 
\bibitem{Keister:1991sb} 
  B.~D.~Keister and W.~N.~Polyzou,
  Adv.\ Nucl.\ Phys.\  {\bf 20}, 225 (1991).  
  
 
\bibitem{GomezRocha:2012hc} 
  M.~Gomez-Rocha,
  Int.\ J.\ Mod.\ Phys.\ A {\bf 27}, 1250163 (2012).
 

 \bibitem{Sokolov:1977ym} 
  S.~N.~Sokolov,
  Theor.\ Math.\ Phys.\  {\bf 36}, 682 (1979)
  [Teor.\ Mat.\ Fiz.\  {\bf 36}, 193 (1978)].
 
 
\bibitem{Coester:1982vt} 
  F.~Coester and W.~N.~Polyzou,
  Phys.\ Rev.\ D {\bf 26}, 1348 (1982).
 
\bibitem{Siegert1937yt} 
  A.~J.~F.~Siegert,
  Phys.\ Rev.\  {\bf 52}, 787 (1937).  
  
\bibitem{Krassnigg:2003gh} 
  A.~Krassnigg, W.~Schweiger and W.~H.~Klink,
  Phys.\ Rev.\ C {\bf 67}, 064003 (2003)
  [nucl-th/0303063].

  
\bibitem{Krassnigg:2004sp} 
  A.~Krassnigg,
  Phys.\ Rev.\ C {\bf 72}, 028201 (2005)
  [nucl-th/0412017].
  
 
\bibitem{Biernat:2014dea} 
  E.~P.~Biernat and W.~Schweiger,
  Phys.\ Rev.\ C {\bf 89}, 055205 (2014)
  [arXiv:1404.2440 [hep-ph]].
 

\bibitem{Biernat:2009my} 
  E.~P.~Biernat, W.~Schweiger, K.~Fuchsberger and W.~H.~Klink,
  Phys.\ Rev.\ C {\bf 79}, 055203 (2009)
  [arXiv:0902.2348 [nucl-th]].
   
 
\bibitem{Biernat:2011mp} 
  E.~P.~Biernat,  PhD Thesis, University of Graz (2011).
  arXiv:1110.3180 [nucl-th].
 

\bibitem{Biernat:2010tp} 
  E.~P.~Biernat, W.~H.~Klink and W.~Schweiger,
  Few Body Syst.\  {\bf 49}, 149 (2011)
  [arXiv:1008.0244 [nucl-th]].
  
  
\bibitem{GomezRocha2012} 
  M.~Gomez-Rocha and W.~Schweiger,
  Phys.\ Rev.\ D {\bf 86}, 053010 (2012)
  [arXiv:1206.1257 [hep-ph]].
  

  \bibitem{GomezRocha:2012hca} 
  M.~Gomez-Rocha and W.~Schweiger,
  Acta Phys.\ Polon.\ Supp.\  {\bf 6}, 365 (2013)
  [arXiv:1211.0868 [hep-ph]].
  
  
\bibitem{Rocha:2010wm} 
  M.~Gomez Rocha and W.~Schweiger,
  Few Body Syst.\  {\bf 50}, 227 (2011)
  [arXiv:1011.0547 [hep-ph]].
 
 
\bibitem{Gomez-Rocha:2013zma} 
  M.~Gomez-Rocha, W.~Schweiger and O.~Senekowitsch,
  arXiv:1311.1936 [hep-ph].
  
  
\bibitem{Senekowitsch2014} 
   O.~Senekowitsch. 
Diploma thesis, Karl-Franzens Universit\"at Graz (2014).

  
\bibitem{Gomez-RochaThesis} 
  M.~G\'omez-Rocha, 
  PhD Thesis, University of Graz (2013).
  arXiv:1306.1248 [hep-ph].
  
  
 \bibitem{Kleinhappel:2013tza} 
  R.~Kleinhappel, W.~Plessas and W.~Schweiger,
  Few Body Syst.\  {\bf 54}, 339 (2013).
  

  \bibitem{Kleinhappel:2012zj} 
  R.~Kleinhappel and W.~Schweiger,
  PoS QNP {\bf 2012}, 076 (2012)
  [arXiv:1206.4213 [hep-ph]].
   
   
\bibitem{Bakamjian:1953kh} 
  B.~Bakamjian and L.~H.~Thomas,
  Phys.\ Rev.\  {\bf 92}, 1300 (1953).
 
   
  
\bibitem{Isgur1989vq} 
  N.~Isgur and M.~B.~Wise,
  Phys.\ Lett.\ B {\bf 232}, 113 (1989).
  
  
\bibitem{Isgur1989ed} 
  N.~Isgur and M.~B.~Wise,
  Phys.\ Lett.\ B {\bf 237}, 527 (1990).
  
  
\bibitem{NeubertHQS} 
  M.~Neubert,
  Phys.\ Rept.\  {\bf 245}, 259 (1994)
  [hep-ph/9306320].
 
 
\bibitem{Keister:2011ie} 
  B.~D.~Keister and W.~N.~Polyzou,
  Phys.\ Rev.\ C {\bf 86}, 014002 (2012)
  [arXiv:1109.6575 [nucl-th]].
 

\bibitem{Carbonell:1998rj} 
  J.~Carbonell, B.~Desplanques, V.~A.~Karmanov and J.~F.~Mathiot,
  Phys.\ Rept.\  {\bf 300}, 215 (1998)
  [nucl-th/9804029].
   
   
   \bibitem{GomezRocha:2011qs} 
  M.~Gomez-Rocha, E.~.P.~Biernat and W.~Schweiger,
  Few Body Syst.\  {\bf 52}, 397 (2012)
  [arXiv:1110.2355 [hep-ph]].
  
\bibitem{Glazek:1989rr} 
  S.~D.~Glazek and M.~Sawicki,
  Phys.\ Rev.\ D {\bf 41}, 2563 (1990).

  
\bibitem{deMelo:2005cy} 
  J.~P.~B.~C.~de Melo, T.~Frederico, E.~Pace, and G.~Salme,
  Phys.\ Rev.\ D {\bf 73}, 074013 (2006)
  [hep-ph/0508001].

  
\bibitem{deMelo:2010sw} 
  J.~P.~B.~C.~de Melo and T.~Frederico,
  Nucl.\ Phys.\ Proc.\ Suppl.\  {\bf 199}, 276 (2010)
  [arXiv:1003.2132 [hep-ph]].
   
  

\bibitem{Simula:2002vm} 
  S.~Simula,
  Phys.\ Rev.\ C {\bf 66}, 035201 (2002)
  [nucl-th/0204015].
 
  
\bibitem{Bakker:2003up} 
  B.~L.~G.~Bakker, H.~-M.~Choi and C.~-R.~Ji,
  Phys.\ Rev.\ D {\bf 67}, 113007 (2003)
  [hep-ph/0303002].

  
\bibitem{Jaus:1999zv} 
  W.~Jaus,
  Phys.\ Rev.\ D {\bf 60}, 054026 (1999).
  
  
\bibitem{Jaus:2002sv} 
  W.~Jaus,
  Phys.\ Rev.\ D {\bf 67}, 094010 (2003)
  [hep-ph/0212098].
 
  
\bibitem{Choi:2011xm} 
  H.~-M.~Choi and C.~-R.~Ji,
  Phys.\ Lett.\ B {\bf 696}, 518 (2011)
  [arXiv:1101.3035 [hep-ph]].
  
  
\bibitem{Choi:2010be} 
  H.~-M.~Choi and C.~-R.~Ji,
  Nucl.\ Phys.\ A {\bf 856}, 95 (2011)
  [arXiv:1007.2502 [hep-ph]].
 
  
\bibitem{Cheng:1996if} 
  H.~-Y.~Cheng, C.~-Y.~Cheung and C.~-W.~Hwang,
  Phys.\ Rev.\ D {\bf 55}, 1559 (1997)
  [hep-ph/9607332].
  
  
\bibitem{Klink:1998zz} 
  W.~H.~Klink,
  Phys.\ Rev.\ C {\bf 58}, 3617 (1998).
  

  \bibitem{Karmanov:1998jp} 
  V.~A.~Karmanov,
  Nucl.\ Phys.\ A {\bf 644}, 165 (1998)
  [nucl-th/9802053].
    
  
\bibitem{Klink:2000pp} 
  W.~H.~Klink,
  Nucl.\ Phys.\ A {\bf 716}, 123 (2003)
  [nucl-th/0012031].
  
  
\bibitem{Polyzou:2010zz} 
  W.~N.~Polyzou,
  Phys.\ Rev.\ C {\bf 82}, 064001 (2010)
  [1008.5222].
  

\bibitem{Wirbel:1985ji} 
  M.~Wirbel, B.~Stech and M.~Bauer,
  Z.\ Phys.\ C {\bf 29}, 637 (1985).

  
\bibitem{Beringer:1900zz} 
  J.~Beringer {\it et al.}  [Particle Data Group Collaboration],
  Phys.\ Rev.\ D {\bf 86}, 010001 (2012).
  


  
  
  
  
\end{thebibliography}
\end{document}